\documentstyle[12pt,aaspp4]{article}  

\def\ea{et al. }
\received{~~}
\accepted{~~}
\journalid{}{}
\articleid{}{}
\begin{document}
\input{psfig.tex}

\title{\bf Multiwavelength Monitoring of the BL Lacertae \\
Object PKS 2155--304 in May 1994. \\
I.  The Ground-Based Campaign}

\author{Joseph E. Pesce\altaffilmark{1}, 
C. Megan Urry\altaffilmark{1},
Laura Maraschi\altaffilmark{2},  
Aldo Treves\altaffilmark{3}, \\
Paola Grandi\altaffilmark{4}, 
Ronald I. Kollgaard\altaffilmark{5}$^{,}$\altaffilmark{6}, 
Elena Pian\altaffilmark{1},
Paul S. Smith\altaffilmark{7}$^{,}$\altaffilmark{8},  \\
Hugh D. Aller\altaffilmark{9},
Margo F. Aller\altaffilmark{9}, 
Aaron J. Barth\altaffilmark{10},  
David A. H. Buckley\altaffilmark{11},  \\
Elvira Covino\altaffilmark{12}, 
Alexei V. Filippenko\altaffilmark{10}, 
Eric J. Hooper\altaffilmark{7}, 
Michael D. Joner\altaffilmark{13}$^{,}$\altaffilmark{14}, \\
Lucyna Kedziora-Chudczer\altaffilmark{15},
David Kilkenny\altaffilmark{11},  
Lewis B. G. Knee\altaffilmark{16}$^{,}$\altaffilmark{17}, \\
Michael Kunkel\altaffilmark{18},  
Andrew C. Layden\altaffilmark{19}$^{,}$\altaffilmark{20}, 
Antonio M\'ario Magalh\~aes\altaffilmark{21}, \\
Fred Marang\altaffilmark{11},  
Vera E. Margoniner\altaffilmark{21}, 
Christopher Palma$^{\rm 5,}$\altaffilmark{22}, \\
Antonio Pereyra\altaffilmark{21}, 
Claudia V. Rodrigues\altaffilmark{21}$^{,}$\altaffilmark{23},
Andries Schutte\altaffilmark{24}$^{,}$\altaffilmark{25}, \\
Michael L. Sitko\altaffilmark{26}, 
Merja Tornikoski\altaffilmark{27},
Johan van der Walt\altaffilmark{28},  \\
Francois van Wyk\altaffilmark{11},
Patricia A. Whitelock\altaffilmark{11}, 
Scott J. Wolk\altaffilmark{14}$^{,}$\altaffilmark{29}
}
\altaffiltext{1}{Space Telescope Science Institute, 3700 San Martin Drive, 
  Baltimore, MD 21218.  The Space Telescope Science Institute is operated by 
  the Association of Universities for Research in Astronomy, Inc., under 
  contract with the National Aeronautics and Space Administration.}
\altaffiltext{2}{Osservatorio Astronomico di Brera, via Brera 28, I-20121 
  Milan, Italy.} 
\altaffiltext{3}{SISSA/ISAS, strada Costiera 11, I-34014 Trieste, Italy.}
\altaffiltext{4}{IAS/CNR, via Enrico Fermi 23, CP67, I-00044 Frascati, Italy.}
\altaffiltext{5}{Department of Astron. and Astrophys., Penn State Univ., 
  University Park, PA 16802.} 
\altaffiltext{6}{Present address: Fermi National Accelerator Laboratory, 
  Box 500, Batavia, IL, 60510.}
\altaffiltext{7}{Steward Observatory, University of Arizona, Tucson, AZ 85721.} 
\altaffiltext{8}{Present address:  NOAO/KPNO, P.O. Box 26732, Tucson, AZ 85726-6732.}
\altaffiltext{9}{Department of Astronomy, University of Michigan, Ann Arbor, 
  MI 48109-1090.} 
\altaffiltext{10}{Department of Astronomy, University of California, 
  Berkeley, CA 94720-3411.} 
\altaffiltext{11}{SAAO, P.O. Box 9, 7935 
  Observatory, Western Cape, South Africa.} 
\altaffiltext{12}{Osservatorio Astronomico di Capodimonte, via Moiariello 16,
  I-80131 Naples, Italy.} 
\altaffiltext{13}{Dept. of Physics and Astronomy, FB, Brigham Young 
  University, Provo, UT 84602.} 
\altaffiltext{14}{Visiting Astronomer, Cerro Tololo Inter-American 
 Observatory, La Serena, Chile. CTIO is operated by Association of 
  Universities for Research in Astronomy, Inc., under contract
  with the National Science Foundation.}
\altaffiltext{15}{Australia Telescope National Facility, P.O. Box 76, Epping
  NSW 2121, Australia.}
\altaffiltext{16}{Swedish-ESO Submillimetre Telescope, ESO, Casilla 19001, 
  Santiago 19, Chile.} 
\altaffiltext{17}{Also with the Onsala Space Observatory, S-43992 Onsala, 
  Sweden.}
\altaffiltext{18}{Max-Planck-Institut f\"ur Astronomie, K\"onigstuhl 17, 
  69117 Heidelberg, Germany.} 
\altaffiltext{19}{Cerro Tololo Inter-American Observatory,
  Casilla 603, La Serena, Chile.}
\altaffiltext{20}{Present address: McMaster University, Dept. of Physics and Astronomy, 
  Hamilton, ON L8S 4M1 Canada.} 
\altaffiltext{21}{Instituto Astronomico e Geof\'isico, Universidade de S\~ao 
  Paulo, Caixa Postal 9638, S\~ao Paulo SP 01065-970, Brazil.} 
\altaffiltext{22}{Present address:  Department of Astronomy, University of 
  Virginia, PO Box 3818, Charlottesville, VA 22903-0818.}
\altaffiltext{23}{Present address:  Instituto Nacional de Pesquisas 
 Espaciais-INPE Divis\~ao de Astrof\'\i sica-DAS, Caixa Postal 515,
 S\~ao Jos\'e dos Campos, SP 12201-970, Brazil.}
\altaffiltext{24}{Department of Physics, University of Zululand, Private 
  Bag X1001, Kwa-Dlangezwa 3886, South Africa.} 
\altaffiltext{25}{Present address:  Siemens Telecommunications, 270 Maggs 
  Street, Waltloo, Pretoria, South Africa.}
\altaffiltext{26}{Department of Physics, University of Cincinnati, Cincinnati, 
  OH 45221-0011.} 
\altaffiltext{27}{Metsahovi Radio Research Station, Metsahovintie 114, 
  FIN-02540 Finland.} 
\altaffiltext{28}{Space Research Unit, Potchefstroom University, 
 Potchefstroom 2520, South Africa.} 
\altaffiltext{29}{SUNY, Stony Brook, New York 11794-2100.}

\begin{abstract}

Optical, near-infrared, and radio observations of the BL Lac object PKS
2155--304 were obtained simultaneously with a continuous UV/EUV/X-ray monitoring
campaign in 1994 May.  Further optical observations were gathered throughout
most of 1994. The radio, millimeter, and near-infrared data show no strong
correlations with the higher energies.  The optical light curves exhibit
flickering of 0.2-0.3 mag on timescales of 1-2 days, superimposed on longer
timescale variations. Rapid variations of $\sim$0.01 mag min$^{-1}$, which,
if real, are the
fastest seen to date for any BL Lac object.  Small (0.2-0.3 mag) increases in
the $V$ and $R$ bands occur simultaneously with a flare seen at higher energies. 
All optical wavebands ({\sl UBVRI\/}) track each other well over the period of
observation with no detectable delay.  For most of the period the average
colors remain relatively constant, although there is a tendency for the colors
(in particular $B-V$) to vary more when the source fades.  
In polarized light, PKS 2155--304 showed strong color dependence (polarization
increases toward the blue, $P_U/P_I = 1.31$) and the highest optical
polarization ($U = 14.3$\%) ever observed for this source. The polarization variations trace
the flares seen in the ultraviolet flux.  For the fastest variability 
timescale observed, we estimate a central black hole mass of $\lesssim1.5 \times
10^9 (\frac{\delta}{10})~M_{\sun}$, consistent with UV and X-ray constraints and smaller than previously
calculated for this object. 

{\em Subject Headings:} BL Lacertae objects: individual 
(PKS 2155--304) --- galaxies: active --- galaxies: photometry --- polarization 
 
\end{abstract}

\section{Introduction}

%
%Determining the continuum production mechanism is critical to understanding the
%central engine in AGN, a fundamental goal in extragalactic astrophysics. The
%emission regions in AGN are too small to image directly with current
%technology; thus, the only tool for probing source structure is the study of
%variability and polarization at multiple wavelengths. 
%

Among active galactic nuclei (AGNs) the blazar class (BL Lacertae objects and
violently variable quasars) is known for rapid variability, high luminosity,
and high level of polarization. The observed properties of blazars are
currently interpreted as nonthermal (synchrotron and inverse Compton) emission
from an inhomogeneous relativistic jet oriented close to the line of sight
(Blandford \& Rees 1978).  Typical jet models (Ghisellini, Maraschi, \& Treves
1985; Marscher \& Gear 1985; K\"onigl 1989) have a large number of free
parameters and are underconstrained by single epoch spectral distributions.
Combining spectral and temporal information greatly constrains the jet physics,
since different models predict different variability as a function of
wavelength. Elucidating the structure of blazar jets through multiwavelength
monitoring and polarization studies is an essential precursor to understanding
their formation and thus the extraction of energy from the central engine. 

At low frequencies (radio-mm-infrared-optical) this technique has already led
to substantial progress: the evolution of radio flares in time and frequency
has been used to deduce the structure of the outer parts of the jet (Hughes,
Aller, \& Aller 1989).  The variations among the radio bands are well
correlated and lags are typically weeks to months.  In some cases, optical
variations precede radio ones by about a year, although only weak correlations
have been established (Bregman \& Hufnagel 1989). Some blazars also exhibit
intraday variability at optical and radio wavelengths (Wagner \& Witzel 1995,
and references therein). Optical polarimetry shows that the synchrotron
continuum completely dominates the emission from most blazars at optical and
ultraviolet wavelengths (Smith \& Sitko 1991). While variations are present at
all frequencies, blazars are generally most variable at the shortest
wavelengths (optical, UV, X-ray). 

The BL Lac object PKS 2155--304 is an excellent candidate for blazar monitoring
because it is both rapidly variable and bright enough that its variability can
be resolved at wavelengths shorter than optical (Edelson et al. 1995); in
particular, PKS 2155--304 is one of only two blazars (the other being Mrk 421)
that can be monitored sufficiently rapidly with the {\it International
Ultraviolet Explorer} ({\it IUE}) satellite. It is also
one of the brightest extragalactic sources detected with the {\it Extreme
Ultraviolet Explorer} satellite ({\it EUVE}; Marshall, Carone, \& Fruscione
1993; Fruscione \ea 1994).  PKS 2155--304 is one of the strongest X-ray
emitters and is a typical X-ray selected BL Lac object. 

High energy $\gamma$-rays from PKS 2155-304 have been detected recently by the
EGRET experiment on board the {\it Compton Gamma-Ray Observatory} ({\it CGRO};
Vestrand, Stacy, \& Sreekumar 1995), confirming that the emission processes in
PKS 2155--304 are similar to those in the many blazars already detected with
{\it CGRO}. Thus, by studying multiwavelength variability in this bright and
highly variable object, we derive information relevant for the whole class,
especially for the ``high-frequency peaked BL Lacs'' (Padovani \& Giommi 1995),
i.e.,  X-ray selected BL Lacs. 

%
%Using the Hubble Space Telescope (HST) Allen \ea (1993) studied the UV
%spectropolarimetry of PKS 2155--304 and found that the UV polarized flux
%is produced by the same mechanism that produces the optical polarization
%and that bother are produced in the same region, confirming earlier
%work by Smith \& Sitko (1991).
%

Attempts at multiwavelength studies of PKS 2155--304 with {\it IUE} and {\it
EXOSAT} (Treves \ea 1989) indicated a correlation of the two wavebands but also
the need for much better sampling. Multiwavelength monitoring of PKS 2155--304
in 1991 November (Smith \ea 1992; Urry \ea 1993; Brinkmann \ea 1994;
Courvoisier \ea 1995; Edelson \ea 1995) produced the best available data for
any blazar. This soft X-ray/UV/optical monitoring of PKS 2155--304 
found the emission at these wavelengths was well correlated, that
there was significant short timescale variability ($<$ 1 day), and that the
X-ray flux led the ultraviolet by a few hours. The tight X-ray/UV correlation
and the overall UV to X-ray spectral shape confirmed the supposition that
synchrotron emission is responsible for the optical-through-X-ray continuum in
this BL Lac object (and presumably in others with similar spectra and
variability), and ruled out conclusively any observed optical/UV continuum from
an accretion disk (as argued also on the basis of polarization studies in the
optical/UV). However, this campaign had sufficient sampling only over a short
period of time (4 days). 

For this reason a second campaign was organized in 1994 May where the intensive
{\it IUE} monitoring was extended to 10 days.  The ultraviolet, extreme
ultraviolet, and X-ray observations, as well as the overall multiwavelength
campaign, are addressed elsewhere (Pian \ea 1996; Marshall \ea 1996; Kii \ea
1996; Urry \ea 1996).  Here we discuss the ground-based
observations during the 1994 May campaign and beyond. 

This paper is organized as follows.  In Section 2 we present the ground-based
observations made from 1994 May through 1994 November. Section 3 includes a
discussion of these data, and conclusions are given in Section 4. 

\section{Multiwavelength Ground-Based Observations}

\subsection{Radio}

The Very Large Array (VLA)\footnote{The National Radio Astronomy Observatory is
operated by Associated Universities, Inc., under cooperative agreement with the
NSF.} was used in a hybrid A/B configuration to monitor the arcsecond core of
PKS 2155--304 on 12 days (1994 May 14 - June 1) at 3 frequencies (8.4, 15.0,
and 22.5 GHz), with 1.5 and 5.0 GHz measurements also taken on four of these
occasions (see Table 1). Standard frequency settings and dual 50 MHz bandwidths
were used.  The uncertainties listed in the Table are the internal errors and
do not include systematic effects, which will alter the overall flux scale (see
below). 

Observations of a few minutes were made at each frequency with similar but not
identical $uv$-coverage on the different days.  Complementary observations
(once per day at each frequency) were also made of 3C\,48 and one or both of
two nearby calibrator sources (2151--304, 2248--325) which we assumed to be
non-variable. Due to poor $uv$ coverage of 3C 48, standard VLA calibration
techniques failed, and we determined the flux densities directly from the raw
$uv$ data.  The absolute flux scale is therefore dependent upon the overall
gain normalization applied to the data during calibration and should be
accurate to 10\% at 22.5 GHz and 5\% at the four lower frequencies (R.\ A.\
Perley, private communication). However, the flux densities obtained from
3C\,48 and the calibrator sources show that the relative flux scale is better
than this, with variations of 1\%, 1\%, 3\%, 3\%, and 4\% noted at 1.5, 5, 8.4,
15, and 22.5 GHz, respectively.  At all frequencies PKS 2155--304 was more
variable than the calibrator sources.  The data collected on May 26 (MJD
9499.05)\footnote{In this paper, MJD is defined as
JD - 2,440,000.} were systematically low for all sources and have been scaled by
assuming that the calibrators are non-variable. 

Data at three frequencies (4.8, 8, and 14.5 GHz) were also obtained with the
University of Michigan Radio Astronomy Observatory (UMRAO) 26 m single-dish
telescope (Table 2). The observational technique and reduction procedures are
discussed by Aller \ea (1985). Typically, each daily observation consists of
a series of on-off measurements over a 30 to 45 minute time period. The flux
scale is based on observations of 3C 461 and the absolute scale of Baars \ea
(1977).  This primary standard, or a nearby secondary flux standard (one of 3C
58, 3C 144, 3C 145, 3C 218, 3C 274, 3C 286, 3C 353, or 3C 405), was routinely
observed every 1.5 to 2 hours to correct for time-dependent variations in the
gain of the instrument.  There is a 5\% uncertainty in the final flux density
scale. 

%
%The main fraction of the data have been acquired as a part of the flux
%monitoring program of the sample of 125 QSOs and BL Lacs in search for the
%intraday variations. These data consist of two epoch of observations carried
%out on the  19th May 94 and 30th of August 94. 
%

Further radio observations were obtained with the Australia Telescope Compact
Array (ATCA; Frater, Brooks, \& Whiteoak 1992), at the Australia Telescope
National Facility, on 1994 May 4-5, May 19-22, and August 30-31. PKS 2155--304
was monitored as part of a program to search for intraday variations in a
sample of quasars and BL Lac objects. The ATCA consists of six 25 m antennas
arranged in an east-west line and observations were done at four wavelengths
(3, 6, 13, and 20 cm). Two slightly different configurations were used
throughout the monitoring, 6A and 6D, with maximum baselines 5939 m and 5878 m,
respectively. The data were taken at 3 and 6 cm simultaneously, then at 13 and
20 cm after rotating the turrets. The correlator was configurated in the
standard way with 32 channels across a bandwidth of 128 MHz for each
wavelength.  During the monitoring program the source was scanned four times
every 24 hours, on average. Each scan lasted one minute with integration times
of 10 seconds. A turret was rotated between two pairs of wavelengths  every
second minute to provide almost simultaneous coverage of the available radio
spectrum, with two orthogonal linear polarizations being measured. 

%
%The source was observed at four wavelengths again but with less frequent turret
%rotation (every 1/2 hour). 
%

The flux density scale was set on the standard primary calibrator used at ATCA,
PKS 1934--638. The changes in phase caused by the receiver, local oscillator,
and atmosphere were calibrated on the nearby point source (the secondary
calibrator), PKS 2149--307. The flux densities given in Table 3 are the
averages over all 13 baselines, with the exception of the two shortest
baselines in order to reduce the influence of the extended structure of PKS
2155--304 and other confusing sources in the field. This is particularly
important at 20 cm where the size of the primary beam is the largest (33 arcmin). 

\subsection{Millimeter}

Observations were made using the 15 m Swedish-ESO Submillimetre Telescope
(SEST)\footnote{The Swedish-ESO Submillimetre Telescope, SEST, is operated
jointly by ESO and the Swedish National Facility for Radio Astronomy, Onsala
Space Observatory at Chalmers University of Technology.}, located on La Silla,
Chile (Booth \ea 1989), with the SEST facility bolometer. PKS 2155--304 was
observed on 1994 May 19 and May 21 at 94 GHz and on 1994 April 24-25 and June
1, 25, and 26 at 90 and 230 GHz (Table 4).  Uranus was used as the primary flux
calibrator and was checked by the secondary calibrator, Jupiter. 

\subsection{Optical and Near-IR Photometry, Polarimetry, and Spectroscopy}

We arranged considerable observational coverage, but bad weather at several
sites prevented the almost continuous level originally planned. The difficulty
of obtaining continuous optical/near-infrared monitoring from the ground was
exacerbated by the fact that the object was $\sim$ 90$^{\circ}$ from the sun in
May, as required for the space-based observations. After May, weather and
sun-angle conditions improved and monitoring observations continued to 1994
November. 

Table 5 lists the 20 optical and near-infrared observers and telescopes
contributing to this campaign. Limited near-infrared data were available during
the middle of the campaign and are given in Table 6. Exposure times were
typically $\sim$40 s for {\sl JHK\/} and $\sim$ 40-160 s for {\sl L\/}.
Observations by M. Kunkel were in the ESO IR system and have been converted to
the SAAO system following Carter (1990). For the optical data (Table 7),
instrumental magnitudes were converted to {\sl UBVRI\/} magnitudes (Johnson
{\sl UBV\/} and Cousins {\sl RI\/} filters) using calibration stars in the
field (Hamuy \& Maza 1989; Smith, Jannuzi, \& Elston 1991). Typical exposure
times were 60-120 s ({\sl U\/}), 20-600 s ({\sl B\/}), 30-300 s ({\sl V\/}),
20-300 s ({\sl R\/}), and 30-120 s ({\sl I\/}), and the errors are in the 
range $\lesssim$0.01 to $\sim$0.08 mag with 0.01 mag being a typical value.   
Some of our observations were obtained at relatively high airmass, and the
differential photometry does not adequately  address the 2nd order extinction
term.  However, this term should account for no more than 0.03 mag. 

Optical polarization measurements of PKS 2155--304 were made between 1994 May
13 and May 21 (MJD 9485-9493) using the Two-Holer polarimeter/photometer (Table
8). The instrument, observational procedures, and data reduction are described
in detail by Smith \ea (1992). An 8 arcsec circular aperture was used for all
of the polarimetry, and typical exposure times were three to eight minutes. 

%
%Prior to 1994 May 18 the Two-Holer was mounted on the University of Minnesota 
%1.5m telescope located on Mt. Lemmon, Arizona.  The observations of 1994 May 
%18--20 (MJD 9491-9493) were taken using the Steward Observatory 1.5m 
%telescope which is also located at the summit of Mt. Lemmon.  
%

Optical polarimetry of PKS 2155--304 was also performed by the group at the
University of S\~ao Paulo (USP) with their CCD Imaging Polarimeter (Table 9).
The instrument was used at the Laboratorio Nacional de Astrofisica (LNA),
Brazopolis, with the LNA 1.60 m and USP 0.61 m telescopes, and is described in
detail by Magalh\~aes et al. (1996). Measurement errors are consistent with
photon noise. Instrumental Stokes Q,U values were converted to the equatorial system
from standard star data obtained on the same night. The instrumental
polarization was measured to be less than 0.03\%; being considerably smaller
than the measured errors, no correction has been applied to the data. 

%
%Basically, a rotating,
%achromatic 51 mm halfwave retarder is followed by a fixed, custom built
%double-calcite prism. The calcite forms two images of each object in the 
%field, separated by 1 mm and with orthogonal polarizations between each 
%other. One linear polarization modulation cycle is covered for every 
%90$^{\circ}$ rotation of the waveplate. 
%
%The CCD exposures through the several filters were taken with the waveplate
%click-stepped through positions 22.5$^{\circ}$ apart. 
%
%[A special purpose FORTRAN routine
%processes data files created within IRAF and calculates from a least
%squares solution the normalized linear polarization Stokes parameters Q
%and U as well as the theoretical (i.e., photon noise) and measurement
%errors.] 
%
%The measurement errors were obtained from the residuals of the observations at
%each position angle 
%
%with regard to the 4-cosine curve
%
%and these are the values quoted in Table 9. They are, as a rule, entirely
%consistent with the photon noise errors. The reader is referred to 
%Magalh\~aes,
%Benedetti, \& Roland (1984) for the pertinent equations. 
%

Optical spectra of PKS 2155--304 were obtained in morning twilight on 1994 June
3 (MJD 9506.9875) with the Kast double spectrograph (Miller \& Stone 1993) at
the Cassegrain focus of the Shane 3 m reflector at Lick Observatory. Reticon 400
$\times$ 1200 pixel CCDs were used in both cameras.  A long slit of width 4
arcsec was oriented along the parallactic angle to minimize differential light
losses produced by atmospheric dispersion. Several different grating and grism
settings were required to cover the entire accessible wavelength range
(3220-9908 \AA ) with a resolution of 8-11 \AA .  The standard stars
BD+26$^{\circ}$2606 (Oke \& Gunn 1983) and Feige 34 (Massey \ea 1988) were used
for flux calibration.  These were also used to eliminate (through division) the
telluric absorption bands in the spectrum of PKS~2155--304. 
The atmospheric seeing during the observations was poor and variable ($\sim$ 3
- 4$^{\prime\prime}$).  Moreover, the extinction correction cannot be fully
trusted because the airmass was high (3.0-3.2).  Thus, although the night was
clear, the derived absolute flux for the final spectrum might be somewhat
erroneous.  The relative flux calibration, on the other hand, should be more
reliable, except perhaps at the near-UV wavelengths. 

\section{Results and Discussion}

\subsection{Radio Results}

The radio fluxes show evidence of variability at the level of a few percent at
all frequencies (see Figure 1), and are better sampled than during the previous
campaign in 1991 November when the flux increased by 20\% in one month (Courvoisier
\ea 1995).  The trend in the high frequency data (22.5, 15, 8.4 GHz) is an
increase of about 10\% from May 14 until around May 24 (MJD 9487 - 9497) when
the flux begins to decline. This trend may also be present in the less-well
sampled 5 GHz data, though probably not at 1.5 GHz. The variability
amplitude appears to
increase with increasing frequency, from $\sim$10\% to $\sim$20\% for the 8.4 to
22.5 GHz data, and the peak at 8.4 and 15 GHz appears to occur simultaneously,
while the 22.5 GHz peak seems to have occurred about five days earlier. 
There is no discernable change in the radio
spectral index, unlike the case in the previous campaign where the radio
spectrum flattened over the period of observation (Courvoisier et al. 1995).
The large 22.5 GHz peak on May 26 (MJD 9499.05) is probably an artifact due to
calibration uncertainties. The Michigan data do not show the same behavior
because of the lack of coverage. 

%
%but also because of the resolution, compared to the VLA data. The latter probes
%the core, while the former measures the more extended regions of the source. 
%

The lower frequency radio data (ATCA) show no variability over the three
periods of observation.  Variations of 20\% would have been observed easily but
are not seen (Figure 2).  PKS 2155--304 does increase in brightness by 20-40\%
from early May to late August at all four wavelengths. Although by a smaller
factor, this corresponds to the general brightening of the source in the
optical over the same period (see below). A marked change in spectral index is
observed for these data, with the spectrum inverting from early May to mid May
and flattening by the last observations in late August. During the SEST
observations the source remained invariant within the errors (Figure 3).
However, variations of $\sim$30\% could be present in the data, comparable to
the radio and optical variations. 

\subsection{Optical and Near-IR Results}
\subsubsection{Photometry and Variability}

The near-infrared flux of PKS 2155--304 increased nearly monotonically by
$\sim$ 0.2-0.3 mag over a period of seven days in all observed bands (Figure
4),  during the ultraviolet flaring period (Pian et al. 1996). The low $L$
magnitude is probably spurious and has much larger errors ($\sim$0.6 mag) than
the other measurements.  The dip seen in $H$ and $J$ may be real, but
instrumental effects cannot be excluded. 

Figure 5 shows the optical ({\sl UBVRI\/}) light curves for PKS 2155--304
during 1994 May. The durations of the X-ray, extreme ultraviolet, and
ultraviolet campaigns ({\it ASCA, EUVE}, and {\it IUE}) are shown at the top
(the middle of each flare is also indicated). While the coverage is sparse,
general trends are the same in all wavebands. The sharp increase of 0.3 mag in
the $V$-band flux, and to a lesser extent in the $R$-band flux, between MJD
9492 and 9494 corresponds to the flare seen in the ultraviolet at the same
period (Urry et al. 1996).  This increase does not seem present in the $B$ band
(unless earlier), and there is no simultaneous data in the $U$ or $I$ bands. 

The entire April - November light curve for PKS 2155--304 is shown in Figure 6.
 There is a general ``flickering'' (mini-flares of $\sim$0.2-0.4 mag in several
days) of the source in all bands throughout this period, superimposed on a
general, slow brightening (0.4-0.7 mag) through September, followed by a 0.4 -
0.7 magnitude drop between the last observations in September (MJD 9608) and
the final observations in November (MJD 9672). Throughout the observation
period, PKS 2155--304 was brighter than the average ($B=13.58$) seen by Pica et
al. (1988) over the period 1979-1986. In early June (MJD $\sim$9500-9520), a
large flare is seen in all observed optical bands. The amplitude of this
feature is 0.3 - 0.4 mag with a rise time of about 10 days and a duration of
about 20 days, although it may not be resolved. 

The largest observed excursions are a drop in the $B$ band of $\sim$ 0.5 mag in
4 days and almost a magnitude in $U$ in about 10 days, both in May, right at
the start of the multiwavelength observing campaign (MJD 9475 - 9484, Figure
5).  The other bands exhibit a drop in magnitude over this period, but of much
smaller amplitudes ($\sim$0.2 mag). There are no UV or X-ray data during this
period; the drop in flux in the 4.8 GHz band is possibly correlated with the
drop in the $U$ band, though this is most likely coincidental. 

The fastest variations are changes of 0.1-0.2 mag in tens of minutes.  An
example of this is the I-band flux at MJD $\sim$9486 (Figure 5), which
increases by 0.18 mag in 13.5 minutes ($\sim$0.01 mag min$^{-1}$!),
corresponding to a doubling time of 75 minutes; a lesser increase is also seen
in the $B$ and $R$ bands. Several such increases are seen in the other bands
over the observation period.  
In fact, if real, these are the fastest optical variations
seen for any BL Lac (by about a factor of five; for OQ 530,  Carini, Miller, \&
Goodrich 1990 observed an increase of 0.06 mag in 20 minutes). A timescale of
about an hour is consistent with the results of a structure analysis of several
nights of photometry by Paltani et al. (1996), who found that the minimum
timescale of variations is shorter than 15 minutes. The other variations we
observe, $\sim$0.01 mag hour$^{-1}$ or several tenths of a magnitude over days,
have been seen before in PKS 2155--304 (Carini \& Miller 1992) and are typical
for these objects (e.g., BL Lac, OJ 287, Miller, Carini, \& Goodrich 1989;
Carini et al. 1992; 0235+164, Rabbette et al. 1996; 0716+714, Wagner et al.
1996).  The blazar 3C 279, the subject of a similar multiwavelength campaign,
was seen to double its $R$-band flux in 10 days (Grandi et al. 1996). 

The most rapid variations observed give us a minimum doubling time or
variability timescale, $t_{\rm D}$ = 75 min.  This, in fact, may not be a
doubling timescale since we have not observed a true doubling of the flux. 
Nonetheless, we can estimate an upper limit to the black hole mass if we
assume that these variations are caused by radiation generated close to a
supermassive black hole (at $R = 3R_{\rm s}$, where $R_{\rm s} = 2GM/c^2$ is
the Schwarzschild radius), and that the emission is isotropic.  An
estimate of the upper limit to the size of the emitting region is $R \approx
\delta ct_{\rm D}$, where $\delta$ is the Doppler factor which takes
relativistic beaming into account; $\delta \sim 10$.

The limit to the mass of the black hole can be estimated by \begin{displaymath}
M_{\rm var} \approx \frac{Rc^2}{6G} \lesssim \frac{\delta c^3t_{\rm D}}{6G}.
\end{displaymath} \noindent For PKS 2155--304 we calculate $M_{\rm var}
\lesssim 1.5 \times 10^9 (\frac{\delta}{10})~M_{\sun}$, consistent with
constraints based on UV and X-ray observations (Morini et al. 1986; 
Urry et al. 1993), and 
considerably smaller than what was found by Carini \& Miller
(1992), taking into account the relativistic beaming term. 

\subsubsection{Colors and Spectral Shape}

In general, the optical light curves of PKS 2155--304 track each other well. No
lags are detected, although because of poor sampling we may be insensitive to
lags of several days in many cases. The $B-V$ and $V-I$ colors of PKS 2155--304
were calculated from simultaneous or nearly simultaneous measurements (the
majority are within one to five minutes, and eight are within 10-40 minutes).
During 1994 May, the $B-V$ colors varied from 0.2 to 0.5 mag, but for most of
the rest of the observation period, they were near the average value of
$\langle B-V \rangle = 0.32 \pm 0.02$ mag.  Except for three points in early
May, the $V-I$ colors were nearly constant at $\langle V-I \rangle = 0.69 \pm
0.01$ mag (Figure 7, top panel). 

The largest color variations occurred when PKS 2155--304 was faint ($V \gtrsim
12.7$). For $V < 12.7$, the standard deviation of the $B-V$ colors is $0.003$
mag while for $V > 12.7$ it is $0.03$ mag (Figure 7, bottom panel). This
is not due to increased measurement errors when the source fades, since
the average errors in the range $V < 12.7$ and $V > 12.7$ are the same.
The colors are constant, except for observations before MJD 9500, at the
start of the campaign, and at MJD 9672, at the end, when the source was
faint ($\langle V \rangle = 12.87 \pm 0.11$ mag compared to 
$\langle V \rangle = 12.55 \pm 0.19$ mag at the other times).  

%
%During these two periods, PKS 2155-304 was redder than the average ($B-V = 0.36
%\pm 0.09$). 
%

The 1994 June 3 (MJD 9506.9875) Lick spectrum of PKS 2155--304 is shown in
Figure 8. The good data cover the wavelength range 4000 - 7500 \AA . Excessive
noise in the region redward of $\sim$ 7600 \AA\ is an artifact of the
high-amplitude interference fringes produced by the CCD; division by flatfields
did not remove them completely, due to flexure of the spectrograph. Several
weak features are visible in the optical region, but these are likely to be
calibration errors; there appear to be no unambiguous absorption or emission
lines to an equivalent width limit of $\sim$1 \AA , and perhaps even
0.5 \AA\ at most locations. Features typically
observed in the spectra of these objects, if present, would be found at the
locations marked (assuming $z = 0.116$; Falomo, Pesce, \& Treves 1993). 
A power law of index $\alpha = -0.71 \pm 0.02$ (where
$F_{\nu} \propto \nu^{\alpha}$) provides a good fit to the spectrum.  This is
identical to the power-law index derived by Courvoisier \ea (1995) from Lick
spectra of PKS 2155--304 obtained on 1991 October 31 and December 14. 

As a further check of the continuum shape, we converted simultaneous magnitudes
(mostly {\sl UBVRI\/}, covering the range $\sim3600-9000$ \AA ) to fluxes using
zero points from Bessell (1979).  We then fit the continuum with a power law (as
above) to get individual spectral slopes, and find an average $\langle
\alpha_{UBVRI} \rangle = -0.76 \pm 0.03$ (Figure 9, top panel).  This is
consistent with what we found from the Lick spectrum presented here, and what
was found in previous studies (Smith \& Sitko 1991; Smith et al. 1992;
Courvoisier et al. 1995; Paltani et al. 1996).  The spectra are slightly
steeper during the period when $V > 12.7$ (Figure 9, bottom panel).

\subsubsection{Polarization}

As with the 1991 November multiwavelength monitoring campaign (Smith \ea 1992),
the optical polarization exhibited strong variability during 1994 May.  The
degree of linear polarization ($P\/$) ranged from $\sim$3\% to $\sim$14\%
(Table 8 and Figure 10, top panel) and the polarization position angle
($\theta\/$) varied from $\sim$100$^\circ$ to $\sim$150$^\circ$ (Figure 10,
bottom panel). Indeed, a change in $\theta\/$ of nearly 25$^\circ$ was observed
between May 15 and May 17 (MJD 9487.96 and 9489.95). 

Broad-band {\sl UBVRI\/} polarimetry acquired on May 19-21 (MJD
9491.96-9493.94) shows the development of strong wavelength-dependent
polarization. Since only $V$-band measurements were made prior to May 19 (MJD
9492), it is impossible to know how $P\/$ and/or $\theta\/$ changed with
wavelength during this period. However, it is apparent that any wavelength
dependence was weak on May 19 ($P_U / P_I = 1.03 \pm 0.19$), while on May 20
(MJD 9493) $P\/$ clearly increases toward the blue ($P_U / P_I = 1.15 \pm
0.08$). Strong wavelength dependence is observed on the following night, with
$P_U / P_I = 1.31 \pm 0.04$. Though the position angle exhibits no trend, the
dependence on wavelength of $P\/$ is among the strongest ever observed for
PKS~2155$-$304, and we note that on May 21 (MJD 9493.9) the $U\/$ polarization
(14.3\%) is the highest optical polarization reported for this object (cf.
Smith \ea 1992). The increases in polarization after May 15 and 19 (MJD 9487.96
and 9491.96) correspond to the two ultraviolet flaring events (Urry \ea 1996). 

Figure 11 shows the polarized $V$-band flux as a function of the $V$-band flux,
ordered chronologically. Except for points 6-8, the optical photometry and
polarization were not strictly simultaneous; there is a difference of about
seven hours between measurements for points 1-4 and one day for point 5. There
are no definite trends, and, in fact, PKS 2155--304 becomes both brighter and
fainter as the polarization increases.  The ultraviolet flaring events occur
after observations 3 and 6. 

%
%In addition to the polarimetry, differential $V\/$ photometric measurements of
%PKS~2155$-$304 were made with Two-Holer on the last three nights (Table 9).
%Stars 2 and 3 of the photometric sequence of Smith \ea (1991) served as
%comparison stars and a 16$^{\prime\prime}$ aperture was utilized. From May
%19--21 (during the development of the strong wavelength dependence in $P\/$)
%the object brightened by more than 0.1 magnitude per day (see Figure 4). 
%

Polarization observations later in the year (Table 9) show a general decrease
from about 10\% in July to around 5\% in October (for the $B$ and $V$ bands, at
least).  At the same time, the object was brightening at all optical bands. It
is interesting to note that the polarization position angle shows no preferred
trend with percent polarization; for these observations it decreases with
decreasing polarization, while in May it both increased and decreased with
increasing polarization.

However, the observed position angles of PKS 2155--304 have been mostly
confined between about 90$^{\circ}$  and 150$^{\circ}$ (Smith et al. 1992;
Allen et al. 1993; Jannuzi, Smith, \& Elston 1993). This has also been the case
for the data collected during the period covered in this paper (Tables 8 and 9,
Figure 10 bottom panel), in contrast to, for example, BL Lac itself (Moore et
al. 1982). 
%
%In the Q-U plane, this translates into a random walk
%motion of the polarization values not around the origin but around an offset
%polarization value. 
%
The preferred polarization orientation for PKS 2155--304 may indicate that the
line of sight to PKS 2155-304 is not as close to the symmetry axis as may be the
case for BL Lac.  These two examples also reflect the general difference
between X-ray-selected BL Lac objects (like PKS 2155--304) and radio-selected
BL Lacs (like BL Lac itself) in that X-ray selected objects have preferred position
angles more often than radio selected ones (Jannuzi, Smith, \& Elston 1994). 

The USP CCD imaging polarimetry also allowed measurements of the foreground
stars in the field of PKS 2155--304.  From $V$ filter images taken on July 2,
we selected seven stars. 
%
%objects with $\sqrt{P^{2} - \sigma^{2}}/\sigma > 3$,
%
A weighted average of the measured Q/I and U/I Stokes parameters yielded $P$ =
(0.31 $\pm$ 0.03)\% at 114\fdg5.  In our fields, star No. 5 of Hamuy \& Maza
(1989) shows  $P$ = (0.27 $\pm$ 0.04)\% at 121\fdg1, in excellent agreement with
the field average.  This corroborates earlier findings (Courvoisier \ea 1995)
that the interstellar polarization towards PKS 2155--304 is negligible. 

\section{Conclusions}

In 1994 May the bright BL Lac object PKS 2155--304 was the subject of a large
multiwavelength campaign.  In this paper, we presented the ground-based radio,
near-infrared, and optical results of the campaign, along with additional
observations made throughout the year to 1994  November. 

The 8.4, 15, and 22.5 GHz data seem to vary together over the observation
period. There is possibly a lag of several days between the 8.4 and 15 GHz data
and those at 22.5 GHz.  When compared to the optical data obtained over the
same period, there is no direct correlation, although the 4.8 GHz data from the
first 10 days of observations may correlate with the optical data, with no
measured lags.  Any correlation between the radio and optical could
be spurious, however, since there are variations on timescales of several days in both
wavebands and many large gaps in coverage. 

The millimeter data are essentially invariant, although variations of
$\sim$30\% are possible within the large errors.  The near-infrared points
exhibit a monotonic increase in brightness over their short observation period.
Both of these data sets show variations comparable to what is seen in the
optical and radio and there are no apparent correlations. The light curves in
the optical bandpasses vary together and show similar short- and long-scale
characteristics throughout the observation period. The fastest variations, of
0.01 mag min$^{-1}$, make PKS 2155--304 the most optically rapidly variable BL
Lac observed to date and are similar to timescales observed in the UV (Pian et
al. 1996).  More typically, variations are $\sim$0.01 mag hour$^{-1}$ or
several tenths of a magnitude over days, which is what is seen for other blazars
(e.g., Carini \& Miller 1992). With a large number of assumptions, we limit the
mass of the central black hole to $M_{\rm var} \lesssim1.5 \times 10^9 
(\frac{\delta}{10})~M_{\sun}$; this is
consistent with the mass determined from UV and X-ray
constraints, and considerably less than what was determined previously in the
optical. 

Smith \ea (1992) and Courvoisier et al. (1995) found the source to have
constant color. Trends in $B-V$ color have been noted before in the sense of
bluer $B-V$ as the source fades (Miller \& McAlister 1983; Carini \& Miller
1992), but during the present campaign the opposite occurred, with slightly
redder colors as the source faded, similar to what was found by Treves et al.
(1989) and Smith \& Sitko
(1991). The effect is small, however. The optical fluxes tracked each other
well, indicating that intensive, multi-band optical monitoring is not necessary
for such campaigns.  Instead, the object can be observed several times per
night in all bands, but intensively in just two or three. 

Polarimetry measurements showed marked color dependence of the polarization
(higher polarization toward the blue), in fact the strongest such dependence
ever observed for PKS 2155--304.  The object was also seen to have the highest
optical polarization observed ($U=14.3$\%), although in the range typical for
X-ray selected BL Lacs (Jannuzi et al. 1994). Also typical for X-ray selected
objects is the preferred position angle of the polarization we observed for PKS
2155--304. 

%
%Optical variability of this sort can be expected (Wagner \& Witzel 1995),
%although the nature of the radio-radio variability and the possible 
%correlation
%between the radio (4.8 GHz) and optical ($U$) bands is not expected 
%(Bregman \&
%Hufnagel 1985).  More thorough, simultaneous coverage in the radio and optical
%is necessary to properly address this point, however. 
%
%
%Of great interest are the optical polarization observations.  Variations
%in the polarization seem to be related to the two flaring events seen
%in the ultraviolet (cf. Urry \ea 1996).  This fact underscores the
%importance of polarization observations during multiwavelength
%campaigns, especially now that monitoring projects in the ultraviolet
%are almost impossible.
%

\acknowledgements

J.E.P., E.P., and C.M.U. would like to acknowledge support from NASA grants
NAG5-1918, NAG5-1034, and NAG5-2499. H.D.A. and M.F.A. acknowledge support from
NSF grant AST-9421979, A.V.F. from NSF grant AST-8957063, E.J.H. from NASA
Grant NGT-51152, R.I.K. and C.P. from NASA LTSA NAGW-2120, and P.S.S. from
NASA Grant NAG5-1630. M.D.J. would like to thank the BYU Department of Physics
and Astronomy for continued support of his research. A.M.M. and C.V.R. received
support from the S\~ao Paulo state funding agency FAPESP through grant No.
94/0033-3. The University of Michigan Radio Astronomy Observatory is supported
by the National Science Foundation and by funds from the University of
Michigan. Sergio Ortolani is thanked for providing data. Some observations were
obtained in the service observing mode from the JKT on La Palma; the help of
Vik Dhillon, Derek Jones, Reynier Peletier, and Keith Tritton and the two
service observers, Phil Rudd and Emilios Harlaftis, is greatly appreciated. The
SAAO CCD data (D. Buckley) were obtained using a focal reducer provided by Dr.
M. Shara (STScI). This research made use of the NASA/IPAC Extragalactic
Database (NED), operated by the Jet Propulsion Laboratory, Caltech, under
contract with NASA, and of NASA's Astrophysics Data System Abstract Service
(ADS).

\clearpage

\begin{table}
\begin{center}
\begin{tabular}{ccccccc}
\multicolumn{7}{c}{{\bf Table 1:}  VLA Radio Data~\tablenotemark{a}} \\ \tableline\tableline
Date     &  JD   &   \multicolumn{5}{c}{Flux (mJy)~\tablenotemark{b}} \\ \cline{3-7}
Observed & (--2,440,000) &  1.5 GHz & 5.0 GHz & 8.4 GHz & 15.0 GHz & 22.5 GHz \\ \tableline
1994 May 14& 9487.04 & $\cdots$   & $\cdots$   & 465$\pm 1$ & 499$\pm 1$ & 488$\pm 3$  \\
~~~~~~May 15 & 9488.13 & 419$\pm 1$ & 509$\pm 1$ & 451$\pm 1$ & 504$\pm 1$ & 476$\pm 2$  \\      
~~~~~~May 16 & 9489.08 & $\cdots$   & $\cdots$   & 465$\pm 1$ & 506$\pm 1$ & 481$\pm 2$  \\
~~~~~~May 17 & 9490.08 & $\cdots$   & $\cdots$   & 463$\pm 1$ & 502$\pm 1$ & 547$\pm 2$  \\
~~~~~~May 18 & 9491.07 & $\cdots$   & $\cdots$   & 466$\pm 1$ & 500$\pm 1$ & 527$\pm 2$  \\
~~~~~~May 21 & 9494.10 & 365$\pm 1$ & 515$\pm 1$ & 488$\pm 1$ & 523$\pm 1$ & 556$\pm 2$  \\
~~~~~~May 22 & 9495.06 & $\cdots$   & $\cdots$   & 498$\pm 1$ & 542$\pm 1$ & 536$\pm 2$  \\
~~~~~~May 23 & 9496.09 & $\cdots$   & $\cdots$   & 508$\pm 1$ & 534$\pm 1$ & 524$\pm 3$  \\
~~~~~~May 24 & 9497.03 & $\cdots$   & $\cdots$   & 509$\pm 1$ & 556$\pm 1$ & 524$\pm 2$  \\
~~~~~~May 26 & 9499.05 & $\cdots$   & $\cdots$   & 481$\pm 1$ & 549$\pm 2$ & 632$\pm 3$  \\
~~~~~~May 28 & 9501.05 & 395$\pm 1$ & 537$\pm 1$ & 486$\pm 2$ & 545$\pm 1$ & 517$\pm 3$  \\
~~~~~~Jun 01 & 9505.04 & 392$\pm 1$ & 514$\pm 1$ & 491$\pm 1$ & 505$\pm 1$ & 473$\pm 2$  \\ \tableline
\end{tabular}
\end{center}
\tablenotetext{\rm a}{VLA data from R. I. Kollgaard and C. Palma.} 
\tablenotetext{\rm b}{Errors shown are the internal errors.  Total 
uncertainties are $\sim$10\%, as in Figure 1.}
\end{table}

\begin{table} 
\begin{center} 
\begin{tabular}{ccccc} 
\multicolumn{5}{c}{{\bf Table 2:}  UMRAO Radio Data~\tablenotemark{a}  } \\ \tableline\tableline 
Date  &  JD   &   \multicolumn{3}{c}{Flux (mJy)~\tablenotemark{b}} \\ \cline{3-5}
Observed & (--2,440,000)& 4.8 GHz & 8.0 GHz & 14.5 GHz \\ \tableline 
1994 Apr 29   & 9471.9708 & 470       & $\cdots$ & $\cdots$     \\ 
~~~~~~May 02   & 9475.0574 & $\cdots$  & $\cdots$ & 450 \\ 
~~~~~~May 03  & 9475.9784 & $\cdots$  & $\cdots$ & 420 \\ 
~~~~~~May 04   & 9477.0253 & $\cdots$  & 490      & $\cdots$     \\
~~~~~~May 05  & 9478.0535 & $\cdots$  & $\cdots$ & 430 \\ 
~~~~~~May 06  & 9478.9718 & $\cdots$  & $\cdots$ & 440 \\ 
~~~~~~May 10  & 9482.9608 & 380      & $\cdots$ & $\cdots$     \\ 
~~~~~~May 19   & 9491.9450 & $\cdots$  & 420     & $\cdots$     \\ 
~~~~~~May 20   & 9493.0255 & $\cdots$  & $\cdots$ & 520 \\
~~~~~~May 27   & 9499.9081 & 490       & $\cdots$ & $\cdots$     \\ 
~~~~~~Jul 31   & 9564.8854 & $\cdots$  & 380      & $\cdots$     \\ 
~~~~~~Sep 15   & 9610.6574 & $\cdots$  & $\cdots$ & 510 \\ 
~~~~~~Sep 17   & 9612.7281 & $\cdots$  & 720      & $\cdots$     \\ 
~~~~~~Sep 21   & 9616.6107 & 540       & $\cdots$ & $\cdots$     \\ 
~~~~~~Dec 04   & 9690.5060 & $\cdots$  & 520      & $\cdots$     \\ \tableline 
\end{tabular} 
\end{center} 
\tablenotetext{\rm a}{Data from the University of Michigan Radio Astronomy 
Observatory (UMRAO) courtesy of M. and H. Aller.} 
\tablenotetext{\rm b}{Typical uncertainties on individual measurements 
are 40, 80, and 20 mJy at 4.8, 8.0, and 14.5 GHz, respectively.}
\end{table}

\begin{table} 
\begin{center} 
\begin{tabular}{cccccc} 
\multicolumn{6}{c}{{\bf Table 3:}  ATCA Centimeter Radio Data~\tablenotemark{a} } \\ \tableline\tableline 
Date  &  JD   &   \multicolumn{4}{c}{Flux (mJy)} \\ \cline{3-6}
Observed & (--2,440,000)& 1.380 GHz & 2.378 GHz & 4.800 GHz & 8.640 GHz \\ 
         &             & (20 cm)   & (13 cm) & (6 cm) & (3 cm) \\ \tableline
1994 May 04  & 9477.1708 & 378.1$\pm$48.9 & 383.1$\pm$26.3 & $\cdots$ & $\cdots$ \\
~~~~~~May 04 & 9477.2007 & $\cdots$  & $\cdots$  & 404.9$\pm$21.7 & 336.5$\pm$19.9\\
~~~~~~May 04 & 9477.2236 & 377.2$\pm$44.8 & 380.0$\pm$24.7 & $\cdots$ & $\cdots$ \\
~~~~~~May 04 & 9477.2576 & $\cdots$  & $\cdots$  & 392.4$\pm$21.8 & 326.5$\pm$18.7 \\
~~~~~~May 04 & 9477.2806 & 369.2$\pm$48.0 & 374.1$\pm$23.4 & $\cdots$ & $\cdots$ \\
~~~~~~May 04 & 9477.3160 & $\cdots$  & $\cdots$  & 394.2$\pm$20.5 & 328.7$\pm$17.9 \\
~~~~~~May 04 & 9477.3368 & 369.3$\pm$86.6 & 376.0$\pm$22.8 & $\cdots$ & $\cdots$ \\
~~~~~~May 04 & 9477.3931 & $\cdots$  & $\cdots$  & 401.2$\pm$20.7 & 331.5$\pm$18.3 \\
~~~~~~May 04 & 9477.4063 & 357.8$\pm$33.2 & 377.7$\pm$24.1 & $\cdots$ & $\cdots$ \\
~~~~~~May 04 & 9477.4236 & $\cdots$  & $\cdots$  & 400.8$\pm$21.1 & 330.0$\pm$18.2 \\
~~~~~~May 04 & 9477.4465 & 373.4$\pm$82.2 & 384.1$\pm$22.9 & $\cdots$ & $\cdots$ \\
~~~~~~May 04 & 9477.4688 & $\cdots$  & $\cdots$  & 401.7$\pm$21.8 & 332.6$\pm$18.5 \\
~~~~~~May 04 & 9477.4917 & 376.2$\pm$64.1 & 386.2$\pm$23.6 & $\cdots$ & $\cdots$ \\
~~~~~~May 05 & 9477.5271 & 368.0$\pm$47.8 & 390.0$\pm$24.7 & $\cdots$ & $\cdots$ \\
~~~~~~May 05 & 9477.5493 & $\cdots$  & $\cdots$  & 406.8$\pm$22.1 & 332.1$\pm$20.0 \\
~~~~~~May 05 & 9477.5722 & 370.9$\pm$47.4 & 392.5$\pm$26.5 & $\cdots$ & $\cdots$ \\
~~~~~~May 05 & 9477.5951 & $\cdots$  & $\cdots$  & 413.3$\pm$24.3 & 337.7$\pm$21.8 \\
~~~~~~May 05 & 9477.6097 & 377.9$\pm$51.2 & 392.7$\pm$29.7 & $\cdots$ & $\cdots$ \\ 
~~~~~~May 19 & 9492.1972 & $\cdots$  & $\cdots$  & 417.5$\pm$30.2 & 406.2$\pm$27.6 \\
~~~~~~May 19 & 9492.1979 & 395.1$\pm$35.0 & 406.9$\pm$32.8 & $\cdots$ & $\cdots$ \\
~~~~~~May 19 & 9492.2764 & $\cdots$  & $\cdots$  & 412.6$\pm$27.9 & 399.4$\pm$26.9 \\
~~~~~~May 19 & 9492.2778 & 377.0$\pm$37.7 & 400.4$\pm$31.6 & $\cdots$ & $\cdots$ \\
~~~~~~May 20 & 9492.5583 & $\cdots$  & $\cdots$  & 422.4$\pm$33.4 & 409.5$\pm$34.0 \\  \tableline
\end{tabular} 
\end{center} 
\end{table} 
\begin{table} 
\begin{center} 
\begin{tabular}{cccccc} 
\multicolumn{6}{c}{{\bf Table 3:}  {\it continued.} } \\ \tableline\tableline 
Date  &  JD   &   \multicolumn{4}{c}{Flux (mJy)} \\ \cline{3-6}
Observed & (--2,440,000)& 1.380 GHz & 2.378 GHz & 4.800 GHz & 8.640 GHz \\ 
         &             & (20 cm)   & (13 cm) & (6 cm) & (3 cm) \\ \tableline
1994 May 20 & 9492.5597 & 378.1$\pm$40.4 & 405.2$\pm$40.0 &$\cdots$ & $\cdots$ \\ 
~~~~~~May 20 & 9493.1368 & $\cdots$  & $\cdots$  & 434.3$\pm$30.0 & 414.5$\pm$33.2 \\
~~~~~~May 20 & 9493.1375 & 392.7$\pm$39.0 & 417.0$\pm$35.2 & $\cdots$ & $\cdots$ \\
~~~~~~May 20 & 9493.1958 & $\cdots$  & $\cdots$  & 429.3$\pm$27.9 & 415.5$\pm$29.2 \\
~~~~~~May 20 & 9493.1965 & 405.5$\pm$41.5 & 422.5$\pm$32.5 & $\cdots$ & $\cdots$ \\
~~~~~~May 20 & 9493.2660 & $\cdots$  & $\cdots$  & 426.4$\pm$27.8 & 415.1$\pm$28.2 \\
~~~~~~May 20 & 9493.2667 & 373.6$\pm$37.5 & 412.2$\pm$32.3 & $\cdots$ & $\cdots$ \\
~~~~~~May 20 & 9493.3708 & $\cdots$  & $\cdots$  & 429.2$\pm$26.6 & 419.7$\pm$28.5 \\
~~~~~~May 20 & 9493.3715 & 360.8$\pm$32.1 & 404.1$\pm$32.0 & $\cdots$ & $\cdots$ \\
~~~~~~May 20 & 9493.4583 & $\cdots$  & $\cdots$  & 431.0$\pm$29.6 & 427.8$\pm$29.5 \\
~~~~~~May 20 & 9493.4597 & 366.2$\pm$40.1 & 404.8$\pm$33.8 & $\cdots$ & $\cdots$ \\
~~~~~~May 21 & 9494.1958 & $\cdots$  & $\cdots$  & 422.9$\pm$28.2 & 421.8$\pm$28.8 \\
~~~~~~May 21 & 9494.1965 & 366.3$\pm$40.4 & 392.1$\pm$33.1 & $\cdots$ & $\cdots$ \\
~~~~~~May 21 & 9494.2708 & $\cdots$  & $\cdots$  & 416.0$\pm$27.0 & 416.3$\pm$28.4 \\
~~~~~~May 21 & 9494.2722 & 361.1$\pm$34.6 & 390.7$\pm$30.6 & $\cdots$ & $\cdots$ \\
~~~~~~May 22 & 9494.5611 & $\cdots$  & $\cdots$  & 427.3$\pm$34.9 & 421.5$\pm$35.7 \\
~~~~~~May 22 & 9494.5618 & 383.1$\pm$43.1 & 395.0$\pm$43.3 & $\cdots$ & $\cdots$ \\
~~~~~~May 22 & 9495.1243 & $\cdots$  & $\cdots$  & 429.4$\pm$31.0 & 434.7$\pm$34.0 \\
~~~~~~May 22 & 9495.1250 & 369.0$\pm$46.6 & 384.2$\pm$36.4 & $\cdots$ & $\cdots$ \\
~~~~~~May 22 & 9495.1826 & $\cdots$  & $\cdots$  & 423.3$\pm$28.8 & 433.8$\pm$29.7 \\
~~~~~~May 22 & 9495.1840 & 367.7$\pm$39.5 & 387.3$\pm$34.8 & $\cdots$ & $\cdots$ \\
~~~~~~May 22 & 9495.2528 & $\cdots$  & $\cdots$  & 421.2$\pm$28.3 & 428.1$\pm$30.2 \\
~~~~~~May 22 & 9495.2542 & 367.1$\pm$32.0 & 381.9$\pm$32.5 & $\cdots$ & $\cdots$ \\ \tableline
\end{tabular} 
\end{center} 
\end{table} 
\begin{table} 
\begin{center} 
\begin{tabular}{cccccc} 
\multicolumn{6}{c}{{\bf Table 3:}  {\it continued.} } \\ \tableline\tableline 
Date  &  JD   &   \multicolumn{4}{c}{Flux (mJy)} \\ \cline{3-6}
Observed & (--2,440,000)& 1.380 GHz & 2.378 GHz & 4.800 GHz & 8.640 GHz \\ 
         &             & (20 cm)   & (13 cm) & (6 cm) & (3 cm) \\ \tableline
1994 Aug 30 & 9594.9340 & 474.9$\pm$35.0 & 467.8$\pm$32.6 & $\cdots$ & $\cdots$ \\ 
~~~~~~Aug 30 & 9594.9354 & $\cdots$  & $\cdots$  & 482.2$\pm$30.9 & 479.7$\pm$35.9 \\
~~~~~~Aug 30 & 9595.0035 & 455.3$\pm$32.1 & 463.8$\pm$32.7 & $\cdots$ & $\cdots$ \\
~~~~~~Aug 30 & 9595.0049 & $\cdots$  & $\cdots$  & 478.1$\pm$33.3 & 477.1$\pm$48.7 \\
~~~~~~Aug 30 & 9595.0889 & 460.0$\pm$32.8 & 461.7$\pm$32.9 & $\cdots$ & $\cdots$ \\
~~~~~~Aug 30 & 9595.0903 & $\cdots$  & $\cdots$  & 473.9$\pm$34.2 & 465.4$\pm$45.3 \\
~~~~~~Aug 30 & 9595.1785 & 480.7$\pm$36.2 & 466.6$\pm$34.1 & $\cdots$ & $\cdots$ \\
~~~~~~Aug 30 & 9595.1799 & $\cdots$  & $\cdots$  & 476.2$\pm$29.6 & 472.2$\pm$34.9 \\
~~~~~~Aug 30 & 9595.2653 & 493.8$\pm$37.6 & 474.7$\pm$36.1 & $\cdots$ & $\cdots$ \\
~~~~~~Aug 30 & 9595.2674 & $\cdots$  & $\cdots$  & 491.4$\pm$30.6 & 484.8$\pm$  32.3 \\
~~~~~~Aug 31 & 9595.8764 & 508.6$\pm$36.8 & 486.2$\pm$33.0 & $\cdots$ & $\cdots$ \\
~~~~~~Aug 31 & 9595.8778 & $\cdots$  & $\cdots$  & 486.3$\pm$30.1 & 484.8$\pm$30.0 \\
~~~~~~Aug 31 & 9595.9813 & 502.8$\pm$34.7 & 475.2$\pm$31.1 & $\cdots$ & $\cdots$ \\
~~~~~~Aug 31 & 9595.9826 & $\cdots$  & $\cdots$  & 476.2$\pm$26.0 & 471.5$\pm$27.0 \\
~~~~~~Aug 31 & 9596.0639 & 496.2$\pm$33.5 & 476.9$\pm$32.1 & $\cdots$ & $\cdots$ \\
~~~~~~Aug 31 & 9596.0653 & $\cdots$  & $\cdots$  & 481.5$\pm$25.9 & 475.9$\pm$27.8 \\
~~~~~~Aug 31 & 9596.1597 & 497.4$\pm$32.5 & 477.9$\pm$32.2 & $\cdots$ & $\cdots$ \\
~~~~~~Aug 31 & 9596.1604 & $\cdots$  & $\cdots$  & 483.2$\pm$28.0 & 479.8$\pm$28.3 \\
~~~~~~Aug 31 & 9596.2556 & 503.0$\pm$43.3 & 476.5$\pm$35.5 & $\cdots$ & $\cdots$ \\
~~~~~~Aug 31 & 9596.2569 & $\cdots$  & $\cdots$  & 487.7$\pm$30.2 & 481.2$\pm$31.5 \\ \tableline
\end{tabular} 
\end{center} 
\tablenotetext{\rm a}{Data from L. Kedziora-Chudczer.}
\end{table}

\begin{table}
\begin{center}
\begin{tabular}{ccccc}
\multicolumn{5}{c}{{\bf Table 4:}  SEST Millimeter Data} \\ \tableline\tableline
Date     &  JD   &   \multicolumn{3}{c}{Flux (mJy)} \\ \cline{3-5}
Observed & (--2,440,000) & 90 GHz\tablenotemark{a}& ~94 GHz\tablenotemark{b} & ~230 GHz\tablenotemark{a}  \\ \tableline
1994 Apr 24     & 9467.139 & 355$\pm 71$  & $\cdots$    & $\cdots$    \\
~~~~~~Apr 25     & 9468.122 & 348$\pm 41$  & $\cdots$    & $\cdots$    \\
~~~~~~Apr 25      & 9468.215 & 417$\pm 57$  & $\cdots$    & $\cdots$    \\
~~~~~~May 19     & 9492.094 & $\cdots$     & 367$\pm 78$ & $\cdots$    \\ 
~~~~~~May 21     & 9494.104 & $\cdots$     & 367$\pm 78$ & $\cdots$    \\
~~~~~~Jun 01     & 9504.778 & 430$\pm 81$  & $\cdots$    & $\cdots$    \\
~~~~~~Jun 25      & 9528.751 & $\cdots$     & $\cdots$    & 310$\pm 24$ \\ 
~~~~~~Jun 25      & 9528.788 & $\cdots$     & $\cdots$    & 330$\pm 25$ \\
~~~~~~Jun 26      & 9529.754 & 450$\pm 117$ & $\cdots$    & $\cdots$    \\ \tableline
\end{tabular}
\end{center}
\tablenotetext{\rm a}{Data from M. Tornikoski.}
\tablenotetext{\rm b}{Data from L.B.G. Knee.}
\end{table}

\begin{table}
\begin{center}
\begin{tabular}{llll}
\multicolumn{4}{c}{{\bf Table 5:}  Optical/Near-IR Observers and Telescopes} \\ \tableline\tableline
Observer & Telescope & Filters & Code~\tablenotemark{a} \\ \tableline
A. Barth     & Lick 3m & $\cdots$ & $\cdots$\\
D. Buckley & SAAO 1.9m & Johnson-Cousins ({\sl UBVR$_{\rm c}$I$_{\rm c}$\/}) & DB \\
E. Covino & ESO 1m &Johnson-Cousins ({\sl UBVR$_{\rm c}$I$_{\rm c}$\/})    & EC \\
A. Filippenko & Lick 3m & $\cdots$ & $\cdots$\\
E. Hooper & Steward 90in & Johnson-Cousins ({\sl BVR$_{\rm c}$I$_{\rm c}$\/})  &EH \\
M. Joner & CTIO 0.9m & Johnson-Cousins ({\sl BVR$_{\rm c}$I$_{\rm c}$\/})  & MJ \\
D. Kilkenny & SAAO 0.5m & Johnson-Cousins ({\sl UBVR$_{\rm c}$I$_{\rm c}$\/})  & DK \\
M. Kunkel & ESO 1m & ESO ({\sl JHKL\/}) & MK \\
A. Layden & KPNO 0.9m & Johnson-Cousins ({\sl BVR$_{\rm c}$I$_{\rm c}$\/})  & ALKP \\
A. Layden & CTIO 0.9m & Johnson-Cousins ({\sl BVR$_{\rm c}$I$_{\rm c}$\/})  & ALCT   \\
M. Magalh\~aes\ & Univ. of S\~ao Paulo 0.61m & Johnson ({\sl BV\/})   & MM \\
F. Marang & SAAO 0.5m & Johnson-Cousins ({\sl UBVR$_{\rm c}$I$_{\rm c}$\/})  & FM \\
S. Ortolani & ESO 1.5m Danish & Johnson ({\sl BV\/})  & SO \\
J. Pesce & La Palma JKT (Service) & Johnson-Cousins ({\sl BVR$_{\rm c}$I$_{\rm c}$\/})  & JEP/JKT \\ 
C. Rodrigues & Univ. of S\~ao Paulo 0.61m & Johnson ({\sl BV\/})   & CR\\
A. Schutte & SAAO 1.9m & SAAO ({\sl JHKL\/}) & AS \\
P. Smith & Steward 1.5m & Johnson-Cousins ({\sl UBVR$_{\rm c}$I$_{\rm c}$\/})  & PSS \\
P. Smith  & U. of Minn. 1.5m & Johnson-Cousins ({\sl UBVR$_{\rm c}$I$_{\rm c}$\/}) & $\cdots$ \\
J.  van der Walt & SAAO 1.9m & SAAO ({\sl JHKL\/}) & JvdW \\
F. van Wyk  & SAAO 0.5m & Johnson-Cousins ({\sl UBVR$_{\rm c}$I$_{\rm c}$\/})  & FvW \\ 
P. Whitelock & SAAO 1.9m & SAAO ({\sl JHKL\/}) & PW \\
S. Wolk & CTIO 0.9m & Johnson-Cousins ({\sl BVR$_{\rm c}$I$_{\rm c}$\/}) & SW \\ \tableline
\end{tabular}
\end{center}
\tablenotetext{\rm a}{Observer codes are used in Tables 6 and 7.}
\end{table}

\begin{table}
\begin{center}
\begin{tabular}{ccccccc}
\multicolumn{7}{c}{{\bf Table 6:}  Near-Infrared Data} \\ \tableline\tableline
Date     & JD    & $J$ & $H$ & $K$ & $L$ & Observer~\tablenotemark{a} \\
Observed & (--2,440,000) &     &     &     &     &                \\ \tableline
1994 May 19  & 9491.68 & 11.51$\pm$0.03 & 10.82$\pm$0.03   
& 10.18$\pm$0.03   & 9.14$\pm$0.05  & PW \\
~~~~~~May 20  & 9492.66 & 11.47$\pm$0.03 & 10.78$\pm$0.03 & 
10.13$\pm$0.03 & 9.06$\pm$0.05 & PW \\
~~~~~~May 24  & 9496.68 & 11.36$\pm$0.03 & 10.6$\pm$0.03  & 
9.98$\pm$0.03 & 8.95$\pm$0.05 & PW \\
~~~~~~May 24  & 9496.95 & 11.51$\pm$0.02 & 10.74$\pm$0.01 & 
9.99 $\pm$0.02 & 9.85$\pm$0.6~~  & MK~\tablenotemark{b}  \\
~~~~~~May 25  & 9497.65 & 11.34$\pm$0.03 & 10.62$\pm$0.03 & 
9.99$\pm$0.03 & 8.88$\pm$0.08 & AS/JvdW \\
~~~~~~May 26  & 9498.66 & 11.30$\pm$0.03 & 10.59$\pm$0.03 & 
9.93$\pm$0.03 & 8.88$\pm$0.08 & AS/JvdW \\ \tableline
\end{tabular}
\end{center}
\tablenotetext{\rm a}{Observers are listed in Table 5.}
\tablenotetext{\rm b}{Magnitudes from MK were originally
from the ESO standard system and have been converted to the SAAO system
(see text).}
\end{table}

\begin{table}
\begin{center}
\begin{tabular}{cccccccc}
\multicolumn{8}{c}{{\bf Table 7:}  Optical Data} \\ \tableline\tableline
Date     & JD    & $U$~\tablenotemark{a} & $B$~\tablenotemark{a} & 
$V$~\tablenotemark{a} & $R$~\tablenotemark{a} & $I$~\tablenotemark{a} & Observer~\tablenotemark{b} \\
Observed & (--2,440,000) &     &     &     &     &     &           \\ \tableline
1994 May 02 & 9474.9173 & 12.33    & 13.21     & 13.02     & 12.76    & 12.31 &EC \\
~~~~~~May 04 & 9476.8737 & 12.59    & 13.29     & 12.98     & 12.67    & 12.26 &EC \\
~~~~~~May 05 & 9477.8953 & 12.65    & 13.29     & 13.03     & 12.73    & 12.33 &EC \\
~~~~~~May 08 & 9480.9778 & $\cdots$ & $\cdots$  & $\cdots$  & 12.83    & $\cdots$ &EH \\
~~~~~~May 08 & 9480.9813 & $\cdots$ & 13.66     & $\cdots$  & $\cdots$ & $\cdots$ &EH \\
~~~~~~May 08 & 9480.9826 & $\cdots$ & 13.67     & $\cdots$  & $\cdots$ & $\cdots$ &EH \\
~~~~~~May 08 & 9480.9861 & $\cdots$ & $\cdots$  & 13.14     & $\cdots$ & $\cdots$ &EH \\
~~~~~~May 08 & 9480.9882 & $\cdots$ & $\cdots$  & $\cdots$  & $\cdots$ & 12.42    &EH \\
~~~~~~May 08 & 9480.9896 & $\cdots$ & $\cdots$  & $\cdots$  & 12.84    & $\cdots$ &EH \\
~~~~~~May 11 & 9483.6832 & 13.17    & $\cdots$  & $\cdots$  & $\cdots$ & $\cdots$ &DB \\
~~~~~~May 11 & 9483.6867 & $\cdots$ & $\cdots$  & 12.73     & $\cdots$ & $\cdots$ &DB \\
~~~~~~May 11 & 9483.6876 & $\cdots$ & $\cdots$  & $\cdots$  & 12.46    & $\cdots$ &DB \\
~~~~~~May 11 & 9483.6884 & $\cdots$ & $\cdots$  & $\cdots$  & $\cdots$ & 12.46    &DB \\
~~~~~~May 12 & 9484.6826 & 13.13    & $\cdots$  & $\cdots$  & $\cdots$ & $\cdots$ &DB \\
~~~~~~May 12 & 9484.6857 & $\cdots$ & $\cdots$  & 13.01     & $\cdots$ & $\cdots$ &DB \\
~~~~~~May 12 & 9484.6870 & $\cdots$ & $\cdots$  & $\cdots$  & 12.66    & $\cdots$ &DB \\
~~~~~~May 13 & 9485.6762 & 12.95    & $\cdots$  & $\cdots$  & $\cdots$ & $\cdots$ &DB \\
~~~~~~May 13 & 9485.6786 & $\cdots$ & 13.28     & $\cdots$  & $\cdots$ & $\cdots$ &DB \\
~~~~~~May 13 & 9485.6794 & $\cdots$ & $\cdots$  & 12.85     & $\cdots$ & $\cdots$ &DB \\
~~~~~~May 13 & 9485.6802 & $\cdots$ & $\cdots$  & $\cdots$  & 12.67    & $\cdots$ &DB \\
~~~~~~May 13 & 9485.6810 & $\cdots$ & $\cdots$  & $\cdots$  & $\cdots$ & 12.32    &DB \\
~~~~~~May 14 & 9486.6609 & 13.12    & $\cdots$  & $\cdots$  & $\cdots$ & $\cdots$ &DB \\
~~~~~~May 14 & 9486.6638 & $\cdots$ & 13.35     & $\cdots$  & $\cdots$ & $\cdots$ &DB \\
~~~~~~May 14 & 9486.6645 & $\cdots$ & $\cdots$  & 12.91     & $\cdots$ & $\cdots$ &DB \\ \tableline
\end{tabular}
\end{center}
\end{table}  

\begin{table}
\begin{center}
\begin{tabular}{cccccccc}
\multicolumn{8}{c}{{\bf Table 7:} {\it - continued.} } \\ \tableline\tableline
Date     & JD    & $U$~\tablenotemark{a} & $B$~\tablenotemark{a} & 
$V$~\tablenotemark{a} & $R$~\tablenotemark{a} & $I$~\tablenotemark{a} & Observer~\tablenotemark{b} \\
Observed & (--2,440,000) &     &     &     &     &     &           \\ \tableline
1994 May 14 & ~9486.6652  & $\cdots$  & $\cdots$  & $\cdots$ & 12.66    & $\cdots$ &DB \\
~~~~~~May 14 & ~9486.6659  & $\cdots$  & $\cdots$  & $\cdots$ & $\cdots$ & 12.34    &DB \\
~~~~~~May 14 & ~9486.6726  & $\cdots$  & 13.22     & $\cdots$ & $\cdots$ & $\cdots$ &DB \\
~~~~~~May 14 & ~9486.6734  & $\cdots$  & $\cdots$  & 12.88    & $\cdots$ & $\cdots$ &DB \\
~~~~~~May 14 & ~9486.6742  & $\cdots$  & $\cdots$  & $\cdots$ & 12.58    & $\cdots$ &DB \\
~~~~~~May 14 & ~9486.6753  & $\cdots$  & $\cdots$  & $\cdots$ & $\cdots$ & 12.16    &DB \\ \
~~~~~~May 15 & ~9487.6542  & $\cdots$  & 13.21     & $\cdots$ & $\cdots$ & $\cdots$ &DB \\
~~~~~~May 15 & ~9487.6575  & $\cdots$  & $\cdots$  & 12.79    & $\cdots$ & $\cdots$ &DB \\
~~~~~~May 16 & ~9488.6347  & $\cdots$  & 13.15     & $\cdots$ & $\cdots$ & $\cdots$ &DB \\
~~~~~~May 16 & ~9488.6382  & $\cdots$  & $\cdots$  & 12.81    & $\cdots$ & $\cdots$ &DB \\
~~~~~~May 16 & ~9488.6595  & $\cdots$  & $\cdots$  & $\cdots$ & 12.54    & $\cdots$ &DB \\
~~~~~~May 16 & ~9488.6623  & $\cdots$  & $\cdots$  & $\cdots$ & $\cdots$ & 12.11    &DB \\
~~~~~~May 18 & *9490.9888 & $\cdots$  & $\cdots$  & $\cdots$ & 12.55    & $\cdots$ &ALKP \\
~~~~~~May 18 & *9490.9919 & $\cdots$  & $\cdots$  & 12.87    & $\cdots$ & $\cdots$ &ALKP \\
~~~~~~May 19 & *9491.9618 & $\cdots$  & $\cdots$  & 12.97    & $\cdots$ & $\cdots$ &PSS \\
~~~~~~May 19 & *9491.9846 & $\cdots$  & $\cdots$  & $\cdots$ & 12.64    & $\cdots$ &ALKP \\
~~~~~~May 19 & *9491.9869 & $\cdots$  & $\cdots$  & 12.96    & $\cdots$ & $\cdots$ &ALKP \\
~~~~~~May 20 & *9492.9271 & $\cdots$  & 13.20     & 12.81    & $\cdots$ & $\cdots$ &SO \\
~~~~~~May 20 & *9492.9597 & $\cdots$  & $\cdots$  & 12.82    & $\cdots$ & $\cdots$ &PSS \\
~~~~~~May 20 & *9492.9817 & $\cdots$  & $\cdots$  & $\cdots$ & 12.47    & $\cdots$ &ALKP \\
~~~~~~May 21 & *9493.9563 & $\cdots$  & $\cdots$  & 12.69    & $\cdots$ & $\cdots$ &PSS \\
~~~~~~May 22 & *9494.7889 & $\cdots$  & 13.20     & $\cdots$ & $\cdots$ & $\cdots$ &MJ \\
~~~~~~May 22 & *9494.7917 & $\cdots$  & $\cdots$  & 12.86    & $\cdots$ & $\cdots$ &MJ \\ \tableline
\end{tabular}
\end{center}
\end{table}  

\begin{table}
\begin{center}
\begin{tabular}{cccccccc}
\multicolumn{8}{c}{{\bf Table 7:}  {\it - continued.} } \\ \tableline\tableline
Date     & JD    & $U$~\tablenotemark{a} & $B$~\tablenotemark{a} & 
$V$~\tablenotemark{a} & $R$~\tablenotemark{a} & $I$~\tablenotemark{a} & Observer~\tablenotemark{b} \\
Observed & (--2,440,000) &     &     &     &     &     &           \\ \tableline
1994 May 22 & *9494.7937 & $\cdots$ & $\cdots$ & $\cdots$ & 12.62    & $\cdots$      &MJ \\
~~~~~~May 22 & *9494.7954 & $\cdots$ & $\cdots$ & $\cdots$ & $\cdots$ & 12.18    &MJ \\
~~~~~~May 22 & *9494.8664 & $\cdots$ & $\cdots$ & 12.90    & $\cdots$ & $\cdots$ &MJ \\
~~~~~~May 22 & *9494.8693 & $\cdots$ & 13.24    & $\cdots$ & $\cdots$ & $\cdots$ &MJ \\
~~~~~~May 25 & *9497.8342 & $\cdots$ & $\cdots$ & $\cdots$ & $\cdots$ & 12.04    &MJ \\
~~~~~~May 25 & *9497.8354 & $\cdots$ & $\cdots$ & $\cdots$ & $\cdots$ & 12.05    &MJ \\
~~~~~~May 25 & *9497.8373 & $\cdots$ & $\cdots$ & $\cdots$ & 12.46    & $\cdots$ &MJ \\
~~~~~~May 25 & *9497.8385 & $\cdots$ & $\cdots$ & $\cdots$ & 12.47    & $\cdots$ &MJ \\
~~~~~~May 25 & *9497.8402 & $\cdots$ & $\cdots$ & 12.76    & $\cdots$ & $\cdots$ &MJ \\
~~~~~~May 25 & *9497.8418 & $\cdots$ & $\cdots$ & 12.81    & $\cdots$ & $\cdots$ &MJ \\
~~~~~~May 25 & *9497.8454 & $\cdots$ & 13.06    & $\cdots$ & $\cdots$ & $\cdots$ &MJ \\ 
~~~~~~May 25 & *9497.8491 & $\cdots$ & 13.06    & $\cdots$ & $\cdots$ & $\cdots$ &MJ \\
~~~~~~May 25 & *9497.8528 & $\cdots$ & 13.06    & $\cdots$ & $\cdots$ & $\cdots$ &MJ \\
~~~~~~May 26 & *9498.8535 & $\cdots$ & $\cdots$ & 12.73    & $\cdots$ & $\cdots$ & MM/CR \\
~~~~~~May 26 & *9498.8603 & $\cdots$ & 13.23    & $\cdots$ & $\cdots$ & $\cdots$ &MM/CR \\
~~~~~~May 26 & *9498.8728 & $\cdots$ & 13.08    & $\cdots$ & $\cdots$ & $\cdots$ &MJ \\
~~~~~~May 26 & *9498.8750 & $\cdots$ & 13.08    & $\cdots$ & $\cdots$ & $\cdots$ &MJ \\
~~~~~~May 26 & *9498.8776 & $\cdots$ & $\cdots$ & 12.76    & $\cdots$ & $\cdots$ &MJ \\
~~~~~~May 26 & *9498.8789 & $\cdots$ & $\cdots$ & 12.76    & $\cdots$ & $\cdots$ &MJ \\
~~~~~~May 26 & *9498.8805 & $\cdots$ & $\cdots$ & $\cdots$ & 12.45    & $\cdots$ &MJ \\
~~~~~~May 26 & *9498.8814 & $\cdots$ & $\cdots$ & $\cdots$ & 12.44    & $\cdots$ &MJ \\
~~~~~~May 26 & *9498.8832 & $\cdots$ & $\cdots$ & $\cdots$ & $\cdots$ & 12.04    &MJ \\
~~~~~~May 26 & *9498.8841 & $\cdots$ & $\cdots$ & $\cdots$ & $\cdots$ & 12.02    &MJ \\ \tableline
\end{tabular}
\end{center}
\end{table}  

\begin{table}
\begin{center}
\begin{tabular}{cccccccc}
\multicolumn{8}{c}{{\bf Table 7:}  {\it - continued.} } \\ \tableline\tableline
Date     & JD    & $U$~\tablenotemark{a} & $B$~\tablenotemark{a} & 
$V$~\tablenotemark{a} & $R$~\tablenotemark{a} & $I$~\tablenotemark{a} & Observer~\tablenotemark{b} \\
Observed & (--2,440,000) &     &     &     &     &     &           \\ \tableline
1994 May 27 & ~9499.9454  & $\cdots$ & $\cdots$ & 12.86    & $\cdots$ & $\cdots$ &SW \\ 
~~~~~~May 27 & 9499.9478  & $\cdots$ & $\cdots$ & $\cdots$ & 12.50    & $\cdots$ &SW \\
~~~~~~May 27 & 9499.9495  & $\cdots$ & $\cdots$ & $\cdots$ & $\cdots$ & 12.14    &SW \\
~~~~~~May 31 & 9503.9392  & $\cdots$ & $\cdots$ & 12.74    & $\cdots$ & $\cdots$ &SW \\
~~~~~~May 31 & 9503.9413  & $\cdots$ & $\cdots$ & $\cdots$ & 12.41    & $\cdots$ &SW \\
~~~~~~Jun 03 & 9506.9386  & $\cdots$ & 12.85    & $\cdots$ & $\cdots$ & $\cdots$ &SW \\
~~~~~~Jun 03 & 9506.9411  & $\cdots$ & $\cdots$ & 12.65    & $\cdots$ & $\cdots$ &SW \\
~~~~~~Jun 03 & 9506.9459  & $\cdots$ & $\cdots$ & $\cdots$ & $\cdots$ & 11.92    &SW \\
~~~~~~Jun 04 & 9507.9301  & $\cdots$ & 12.94    & $\cdots$ & $\cdots$ & $\cdots$ &SW \\
~~~~~~Jun 04 & 9507.9325  & $\cdots$ & $\cdots$ & 12.72    & $\cdots$ & $\cdots$ &SW \\
~~~~~~Jun 12 & 9515.5978  & 12.15    & 12.85    & 12.55    & 12.25    & 11.86    &DK/FM/FvW \\
~~~~~~Jun 14 & 9517.6211  & 12.24    & 12.94    & 12.63    & 12.32    & 11.92    &DK/FM/FvW \\            
~~~~~~Jun 15 & 9518.6302  & 12.32    & 13.01    & 12.68    & 12.38    & 11.98    &DK/FM/FvW \\         
~~~~~~Jun 16 & 9519.6477  & 12.38    & 13.07    & 12.74    & 12.44    & 12.08    &DK/FM/FvW \\        
~~~~~~Jun 22 & 9525.6015  & 12.45    & 13.11    & 12.88    & 12.58    & 12.17    &DK/FM/FvW \\               
~~~~~~Jun 26 & 9529.9427  & $\cdots$ & $\cdots$ & $\cdots$ & $\cdots$ & 12.15    &ALCT \\
~~~~~~Jun 26 & 9529.9438  & $\cdots$ & $\cdots$ & 12.86    & $\cdots$ & $\cdots$ &ALCT \\
~~~~~~Jun 26 & 9529.9452  & $\cdots$ & 13.10    & $\cdots$ & $\cdots$ & $\cdots$ &ALCT \\
~~~~~~Jun 27 & 9530.8505  & $\cdots$ & 13.06    & $\cdots$ & $\cdots$ & $\cdots$ &ALCT \\
~~~~~~Jun 27 & 9530.8538  & $\cdots$ & $\cdots$ & 12.80    & $\cdots$ & $\cdots$ &ALCT \\
~~~~~~Jun 27 & 9530.8554  & $\cdots$ & $\cdots$ & $\cdots$ & $\cdots$ & 12.04    &ALCT \\
~~~~~~Jun 27 & 9530.8616  & $\cdots$ & $\cdots$ & $\cdots$ & $\cdots$ & 12.10    &ALCT \\
~~~~~~Jul 01 & 9534.6218  & 12.31    & 13.02    & 12.71    & 12.42    & 12.03    &DK/FM/FvW \\
~~~~~~Jul 02 & 9535.6036  & 12.36    & 13.05    & 12.73    & 12.43    & 12.03    &DK/FM/FvW \\ \tableline
\end{tabular}
\end{center}
\end{table}  

\begin{table}
\begin{center}
\begin{tabular}{cccccccc}
\multicolumn{8}{c}{{\bf Table 7:}  {\it - continued.} } \\ \tableline\tableline
Date     & JD    & $U$~\tablenotemark{a} & $B$~\tablenotemark{a} & 
$V$~\tablenotemark{a} & $R$~\tablenotemark{a} & $I$~\tablenotemark{a} & Observer~\tablenotemark{b} \\
Observed & (--2,440,000) &     &     &     &     &     &           \\  \tableline
1994 Jul 03 & 9536.6211  & 12.35    & 13.05    & 12.74    & 12.44    & 12.06    &DK/FM/FvW \\ 
~~~~~~Jul 04 & 9537.5886   & 12.26 & 12.97 & 12.68 & 12.37 & 12.00 &DK/FM/FvW \\ 
~~~~~~Jul 05 & 9538.5756   & 12.45 & 13.14 & 12.83 & 12.51 & 12.10 &DK/FM/FvW \\
~~~~~~Jul 06 & 9539.5865   & 12.45 & 13.15 & 12.82 & 12.52 & 12.12 &DK/FM/FvW \\
~~~~~~Jul 07 & 9540.5710   & 12.44 & 13.14 & 12.82 & 12.52 & 12.11 &DK/FM/FvW \\
~~~~~~Jul 08 & 9541.5812   & 12.43 & 13.12 & 12.80 & 12.49 & 12.07 &DK/FM/FvW \\
~~~~~~Jul 09 & 9542.6156   & 12.39 & 13.10 & 12.79 & 12.49 & 12.10 &DK/FM/FvW \\
~~~~~~Jul 12 & 9545.5740   & 12.34 & 13.04 & 12.74 & 12.44 & 12.04 &DK/FM/FvW \\
~~~~~~Jul 13 & 9546.5510   & 12.38 & 13.09 & 12.78 & 12.48 & 12.08 &DK/FM/FvW \\
~~~~~~Jul 14 & 9547.5769   & 12.41 & 13.10 & 12.79 & 12.49 & 12.10 &DK/FM/FvW \\
~~~~~~Jul 20 & 9553.6226   & $\cdots$  & 13.02    & $\cdots$ & $\cdots$ & $\cdots$ &JEP/JKT \\
~~~~~~Jul 20 & 9553.6290   & $\cdots$  & 13.02    & $\cdots$ & $\cdots$ & $\cdots$ &JEP/JKT \\
~~~~~~Jul 20 & 9553.6341   & $\cdots$  & 13.02    & $\cdots$ & $\cdots$ & $\cdots$ &JEP/JKT \\
~~~~~~Jul 20 & 9553.6390   & $\cdots$  & $\cdots$ & 12.67    & $\cdots$ & $\cdots$ &JEP/JKT \\   
~~~~~~Jul 20 & 9553.6437   & $\cdots$  & $\cdots$ & 12.66    & $\cdots$ & $\cdots$ &JEP/JKT \\
~~~~~~Jul 20 & 9553.6520   & $\cdots$  & $\cdots$ & $\cdots$ & 12.35    & $\cdots$ &JEP/JKT \\
~~~~~~Jul 20 & 9553.6555   & $\cdots$  & $\cdots$ & $\cdots$ & 12.34    & $\cdots$ &JEP/JKT \\
~~~~~~Jul 20 & 9553.6618   & $\cdots$  & $\cdots$ & $\cdots$ & $\cdots$ & 11.93    &JEP/JKT \\
~~~~~~Jul 20 & 9553.6682   & $\cdots$  & $\cdots$ & $\cdots$ & $\cdots$ & 11.94    &JEP/JKT \\
~~~~~~Jul 28 & 9561.5073   & 12.14     & 12.84    & 12.52    & 12.22    & 11.80    &DK/FM/FvW \\
~~~~~~Jul 29 & 9562.5073   & 12.13     & 12.82    & 12.53    & 12.23    & 11.83    &DK/FM/FvW \\
~~~~~~Jul 30 & 9563.5061   & 12.22     & 12.92    & 12.62    & 12.31    & 11.91    &DK/FM/FvW \\
~~~~~~Aug 01 & 9566.4992   & 12.19     & 12.89    & 12.58    & 12.28    & 11.91    &DK/FM/FvW \\
~~~~~~Aug 02 & 9566.5353   & $\cdots$  & 12.93    & $\cdots$ & $\cdots$ & $\cdots$ &JEP/JKT \\ \tableline
\end{tabular}
\end{center}
\end{table}  

\begin{table}
\begin{center}
\begin{tabular}{cccccccc}
\multicolumn{8}{c}{{\bf Table 7:}  {\it - continued.} } \\ \tableline\tableline
Date     & JD    & $U$~\tablenotemark{a} & $B$~\tablenotemark{a} & 
$V$~\tablenotemark{a} & $R$~\tablenotemark{a} & $I$~\tablenotemark{a} & Observer~\tablenotemark{b} \\
Observed & (--2,440,000) &     &     &     &     &     &           \\ \tableline
1994 Aug 02 & 9566.5425   & $\cdots$  & 12.96    & $\cdots$ & $\cdots$ & $\cdots$ &JEP/JKT \\ 
~~~~~~Aug 02 & 9566.5491 & $\cdots$ & $\cdots$ & 12.60    & $\cdots$ & $\cdots$ &JEP/JKT \\
~~~~~~Aug 02 & 9566.5535 & $\cdots$ & $\cdots$ & 12.62    & $\cdots$ & $\cdots$ &JEP/JKT \\
~~~~~~Aug 02 & 9566.5561 & $\cdots$ & $\cdots$ & $\cdots$ & 12.30    & $\cdots$ &JEP/JKT \\
~~~~~~Aug 02 & 9566.5604 & $\cdots$ & $\cdots$ & $\cdots$ & 12.29    & $\cdots$ &JEP/JKT \\
~~~~~~Aug 02 & 9566.5627 & $\cdots$ & $\cdots$ & $\cdots$ & $\cdots$ & 11.90    &JEP/JKT \\
~~~~~~Aug 02 & 9566.5668 & $\cdots$ & $\cdots$ & $\cdots$ & $\cdots$ & 11.89    &JEP/JKT \\
~~~~~~Aug 10 & 9575.4039 & 12.01 & 12.72 & 12.44 & 12.15 & 11.78 &DK/FM/FvW \\
~~~~~~Aug 10 & 9575.4813 & 12.00 & 12.71 & 12.42 & 12.13 & 11.75 &DK/FM/FvW \\
~~~~~~Aug 11 & 9576.4048 & 12.12 & 12.82 & 12.51 & 12.22 & 11.82 &DK/FM/FvW \\
~~~~~~Aug 11 & 9576.4830 & 12.12 & 12.82 & 12.51 & 12.21 & 11.82 &DK/FM/FvW \\ 
~~~~~~Aug 12 & 9576.5704 & 12.12 & 12.82 & 12.51 & 12.22 & 11.81 &DK/FM/FvW \\ 
~~~~~~Aug 15 & 9580.4852 & 11.96 & 12.65 & 12.36 & 12.07 & 11.68 &DK/FM/FvW \\
~~~~~~Aug 16 & 9580.5576 & 11.97 & 12.67 & 12.36 & 12.07 & 11.66 &DK/FM/FvW \\
~~~~~~Aug 16 & 9581.4061 & 11.92 & 12.62 & 12.32 & 12.04 & 11.67 &DK/FM/FvW \\
~~~~~~Aug 16 & 9581.4547 & 11.90 & 12.60 & 12.31 & 12.01 & 11.63 &DK/FM/FvW \\
~~~~~~Aug 17 & 9581.5542 & 11.88 & 12.55 & 12.28 & 12.00 & 11.66 &DK/FM/FvW \\
~~~~~~Aug 17 & 9582.3906 & 11.93 & 12.61 & 12.32 & 12.02 & 11.62 &DK/FM/FvW \\
~~~~~~Aug 17 & 9582.4599 & 11.90 & 12.61 & 12.30 & 12.02 & 11.62 &DK/FM/FvW \\
~~~~~~Aug 18 & 9582.5257 & 11.90 & 12.61 & 12.31 & 12.02 & 11.65 &DK/FM/FvW \\
~~~~~~Aug 21 & 9586.3861 & 12.06 & 12.70 & 12.40 & 12.14 & 11.72 &DK/FM/FvW \\
~~~~~~Aug 21 & 9586.4330 & 12.04 & 12.76 & 12.44 & 12.13 & 11.75 &DK/FM/FvW \\
~~~~~~Aug 22 & 9586.5211 & 12.07 & 12.81 & 12.49 & 12.16 & 11.79 &DK/FM/FvW \\
~~~~~~Aug 22 & 9587.3876 & 11.99 & 12.72 & 12.39 & 12.09 & 11.66 &DK/FM/FvW \\ \tableline
\end{tabular}
\end{center}
\end{table}  

\begin{table}
\begin{center}
\begin{tabular}{cccccccc}
\multicolumn{8}{c}{{\bf Table 7:}  {\it - continued.} } \\ \tableline\tableline
Date     & JD    & $U$~\tablenotemark{a} & $B$~\tablenotemark{a} & 
$V$~\tablenotemark{a} & $R$~\tablenotemark{a} & $I$~\tablenotemark{a} & Observer~\tablenotemark{b} \\
Observed & (--2,440,000) &     &     &     &     &     &           \\ \tableline
1994 Aug 22 & 9587.4385 & 11.99 & 12.73 & 12.42 & 12.09 & 11.69 &DK/FM/FvW \\ 
~~~~~~Aug 23 & 9587.5223 & 12.03 & 12.76 & 12.46 & 12.09 & 11.77 &DK/FM/FvW \\
~~~~~~Aug 25 & 9590.4331 & 11.99 & 12.67 & 12.34 & 12.02 & 11.61 &DK/FM/FvW \\
~~~~~~Aug 29 & 9594.4065   & 11.98 & 12.68 & 12.36 & 12.04 & 11.65 &DK/FM/FvW \\
~~~~~~Sep 02 & 9597.5145   & 11.91 & 12.61 & 12.31 & 12.01 & 11.59 &DK/FM/FvW \\
~~~~~~Sep 02 & 9598.4908   & 11.99 & 12.67 & 12.36 & 12.06 & 11.65 &DK/FM/FvW \\
~~~~~~Sep 05 & 9601.4694   & 12.03 & 12.71 & 12.36 & 12.05 & 11.64 &DK/FM/FvW \\
~~~~~~Sep 08 & 9604.3906   & 11.90 & 12.59 & 12.27 & 11.96 & 11.55 &DK/FM/FvW \\
~~~~~~Sep 11 & 9607.3721   & 12.00 & 12.70 & 12.39 & 12.09 & 11.71 &DK/FM/FvW \\
~~~~~~Sep 12 & 9608.3965   & 11.87 & 12.57 & 12.28 & 11.98 & 11.59 &DK/FM/FvW \\
~~~~~~Nov 16 & 9672.6097 & $\cdots$  & $\cdots$ & $\cdots$ & 12.44    & $\cdots$ &EH \\
~~~~~~Nov 16 & 9672.6167 & $\cdots$  & 13.24    & $\cdots$ & $\cdots$ & $\cdots$ &EH \\
~~~~~~Nov 16 & 9672.6285 & $\cdots$  & $\cdots$ & 12.74    & $\cdots$ & $\cdots$ &EH \\
~~~~~~Nov 16 & 9672.6403 & $\cdots$  & $\cdots$ & $\cdots$ & $\cdots$ & 12.01    &EH \\ \tableline
\end{tabular}
\end{center}
\tablenotetext{\rm a}{For all passbands, uncertainties are typically 0.01 mag.}
\tablenotetext{\rm b}{Observers are listed in Table 5.}
\tablenotetext{*}{Simultaneous space-based data available on these dates.}
\end{table}  

\clearpage

\begin{table}
\begin{center}
\begin{tabular}{ccccc}
\multicolumn{5}{c}{{\bf Table 8:}  Two-Holer Polarization Data~\tablenotemark{a}} \\ \tableline\tableline
Date         & JD      & Filter & P  & PA  \\
Observed     & (--2,440,000) && (\%)  & ($^{\circ}$)       \\ \tableline
1994 May 19  & 9491.95 & $U$ & ~5.53$\pm$0.94 & 157.1$\pm$4.8 \\
~~~~~~May 20 & 9492.95 & $U$ & ~8.84$\pm$0.48 & 137.9$\pm$1.6 \\
~~~~~~May 21 & 9493.95 & $U$ & 14.25$\pm$0.35 & 133.0$\pm$0.7 \\ \tableline
~~~~~~May 19 & 9491.96 & $B$ & ~6.23$\pm$0.37 & 151.1$\pm$1.7 \\
~~~~~~May 20 & 9492.94 & $B$ & ~8.76$\pm$0.35 & 136.4$\pm$1.1 \\
~~~~~~May 21 & 9493.94 & $B$ & 13.19$\pm$0.30 & 131.8$\pm$0.6 \\ \tableline
~~~~~~May 13 & 9485.97 & $V$ & ~7.57$\pm$0.67 & ~96.5$\pm$2.5 \\ 
~~~~~~May 14 & 9486.96 & $V$ & ~7.52$\pm$0.81 & 101.7$\pm$3.1 \\
~~~~~~May 15 & 9487.96 & $V$ & ~2.88$\pm$0.29 & ~98.0$\pm$2.9 \\
~~~~~~May 16 & 9488.96 & $V$ & ~4.99$\pm$0.22 & 108.9$\pm$1.3 \\
~~~~~~May 17 & 9489.95 & $V$ & ~6.54$\pm$0.36 & 132.1$\pm$1.6 \\
~~~~~~May 19 & 9491.96 & $V$ & ~5.92$\pm$0.28 & 150.1$\pm$1.3 \\
~~~~~~May 20 & 9492.96 & $V$ & ~8.02$\pm$0.30 & 135.3$\pm$1.1 \\
~~~~~~May 21 & 9493.95 & $V$ & 12.31$\pm$0.21 & 131.5$\pm$0.5 \\ \tableline
~~~~~~May 19 & 9491.95 & $R$ & ~5.68$\pm$0.29 & 151.2$\pm$1.4 \\
~~~~~~May 20 & 9492.94 & $R$ & ~7.54$\pm$0.33 & 134.9$\pm$1.2 \\
~~~~~~May 21 & 9493.93 & $R$ & 11.90$\pm$0.21 & 132.0$\pm$0.5 \\ \tableline
~~~~~~May 19 & 9491.95 & $I$ & ~5.38$\pm$0.30 & 150.8$\pm$1.6 \\
~~~~~~May 20 & 9492.95 & $I$ & ~7.70$\pm$0.36 & 135.5$\pm$1.3 \\
~~~~~~May 21 & 9493.94 & $I$ & 10.89$\pm$0.24 & 132.6$\pm$0.6 \\ \tableline
\end{tabular}
\end{center}
\tablenotetext{\rm a}{Observations by P. Smith with the 
Univ. of Minnesota 1.5m (1994 May 13-17) and the Steward Observatory 1.5m 
(1994 May 19-21) telescopes, both on Mt. Lemmon, Arizona.}
\end{table}

\begin{table}
\begin{center}
\begin{tabular}{ccccc}
\multicolumn{5}{c}{{\bf Table 9:}  LNA and USP Polarization Data~\tablenotemark{a}}  \\ \tableline\tableline
Date         & JD      & Filter & P      & PA  \\
Observed     & (--2,440,000) &        & (\%)  & ($^{\circ}$)  \\ \tableline
1994 Jul 05  & 9538.83 & $B$ & 10.04$\pm$0.14 & 114.1$\pm$0.4\\
~~~~~~Oct 13 & 9638.53 & $B$ & ~5.16$\pm$0.06 & ~96.8$\pm$0.3\\ \tableline
~~~~~~Jul 02 & 9535.80 & $V$ & 11.29$\pm$0.15 & 105.0$\pm$0.4\\
~~~~~~Jul 21 & 9554.83 & $V$ & ~8.12$\pm$0.15 & ~94.7$\pm$0.5\\
~~~~~~Sep 01 & 9596.79 & $V$ & ~5.36$\pm$0.04 & ~95.1$\pm$0.2\\
~~~~~~Oct 13 & 9638.57 & $V$ & ~5.36$\pm$0.07 & ~95.5$\pm$0.4\\ \tableline
~~~~~~Jul 02 & 9535.85 & $R$ & 10.28$\pm$0.05 & 101.6$\pm$0.1\\
~~~~~~Jul 21 & 9554.79 & $R$ & ~7.33$\pm$0.10 & ~94.1$\pm$0.4 \\ \tableline
~~~~~~Jul 05 & 9538.71 & $I$ & ~9.20$\pm$0.09 & 118.9$\pm$0.3\\
~~~~~~Jul 25 & 9558.70 & $I$ & ~7.38$\pm$0.07 & 101.4$\pm$0.3 \\ \tableline
~~~~~~Jul 03 & 9536.78 & none& 10.22$\pm$0.08 & 104.3$\pm$0.2\\ \tableline
\end{tabular}
\end{center}
\tablenotetext{\rm a}{Observations by A. Magalh\~aes\, V. Margoniner, 
A. Pereyra,
and C. Rodrigues with the LNA 1.60m and USP 0.61m telescopes.}
\end{table}

\clearpage

\centerline{\psfig{figure=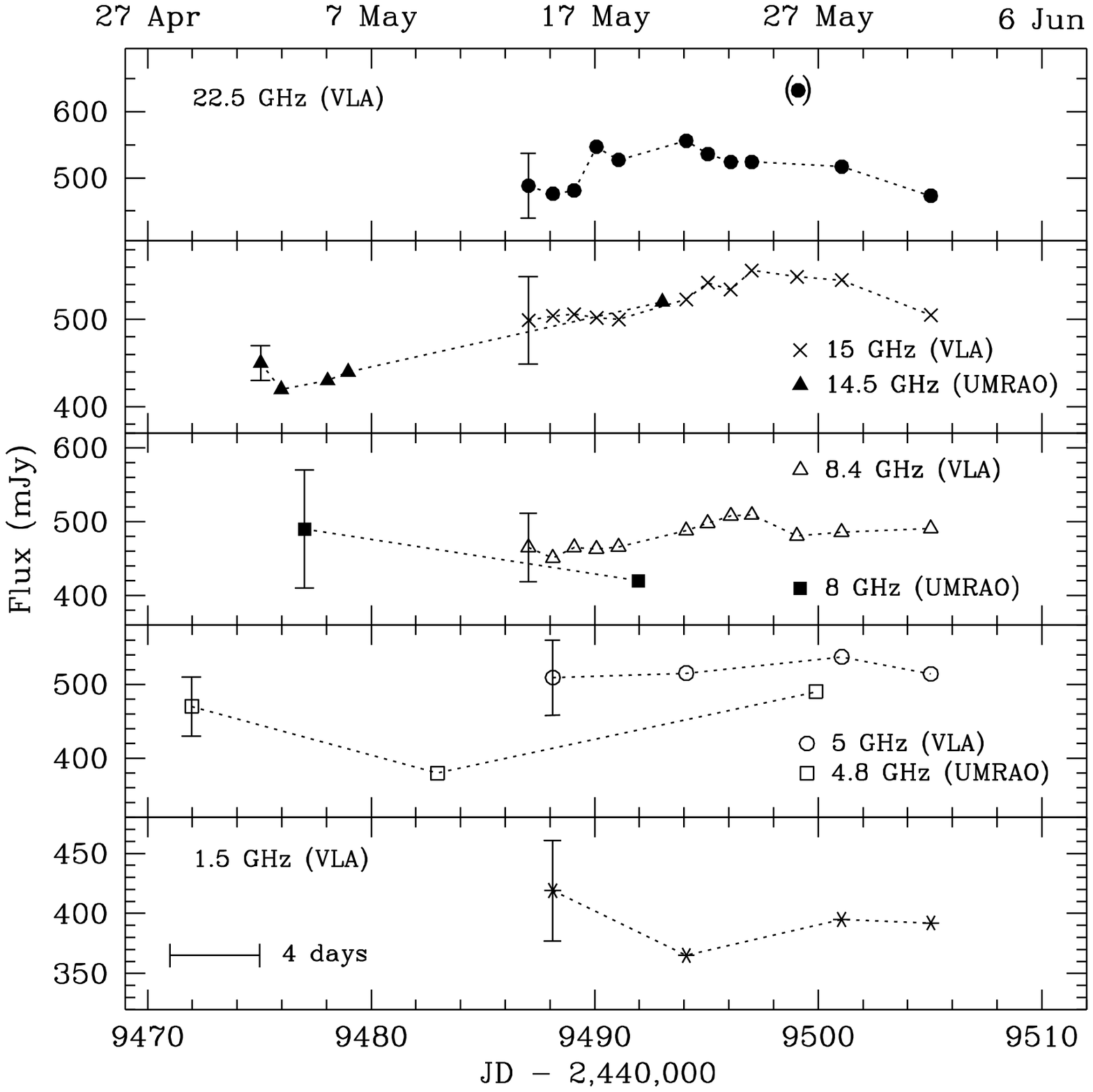,height=5in }}
\figcaption[fig1.ps]{The radio light curves from 1994 April to June. 
The 22.5, 15, and 8.4 GHz data show a slight (10\% - 20\%) increase in flux 
and then a decrease by the same amount over the observation period.
The 14.5 GHz data increase by the same amount, while the other bands
are basically invariant.  Uncertainties are shown on the first
point of each light curve and are $\pm$10\% for the VLA data and $\pm$40, 80,
and 20 mJy for the UMRAO 4.8, 8.0, and 14.5 GHz data, respectively. The large
22.5 GHz peak on May 26 (MJD 9499.05) is probably an artifact.
The lines have been added to guide
the eye only.} 

\centerline{\psfig{figure=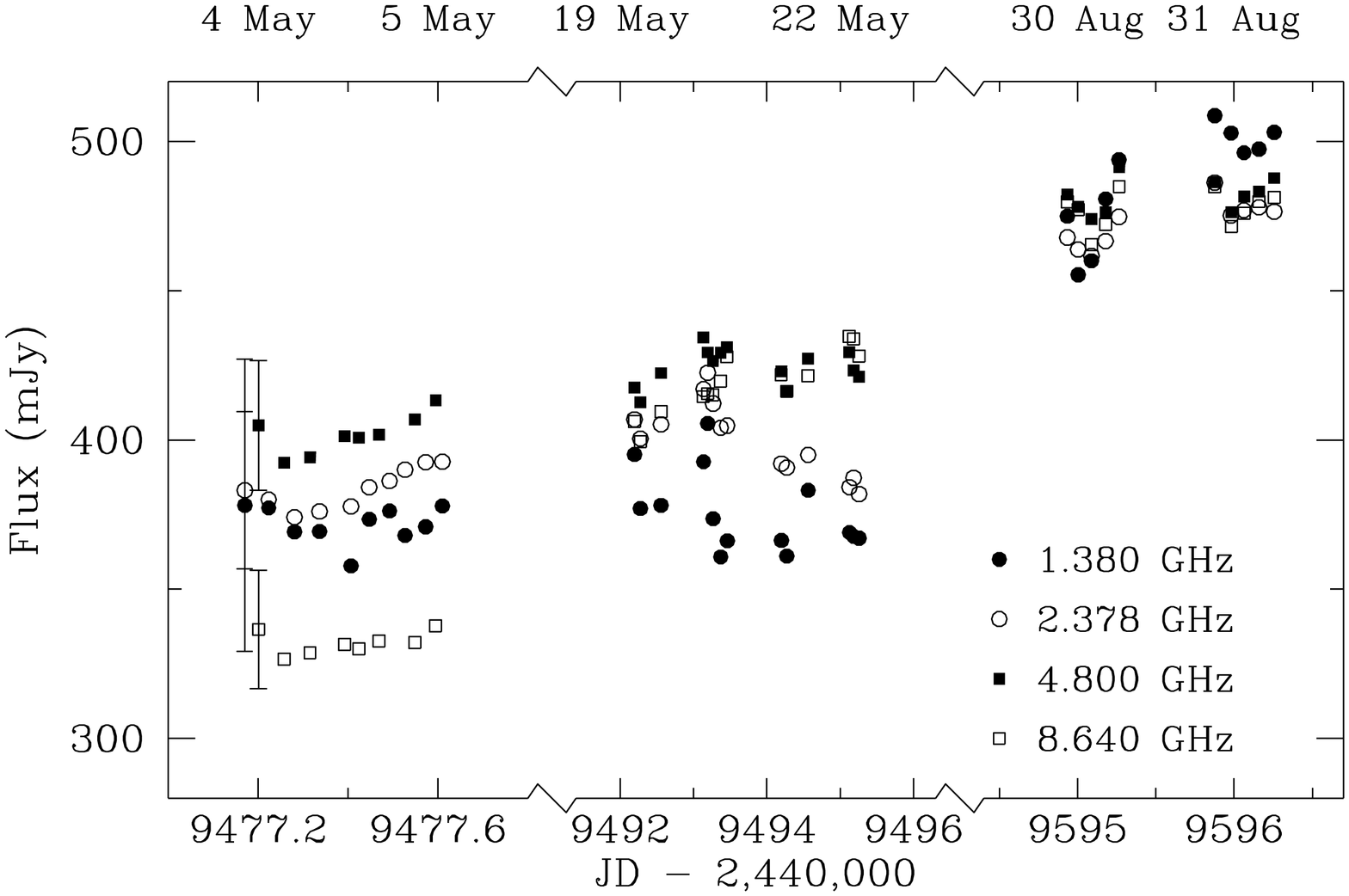,height=5in,angle=270 }}
\figcaption[fig2.ps]{The centimeter light curves from ATCA.  The source
brightens slightly (20-40\%) over the observation period (May - August) 
and there is a strong change in the spectral index between early and mid
May and again between mid May and late August.}

\centerline{\psfig{figure=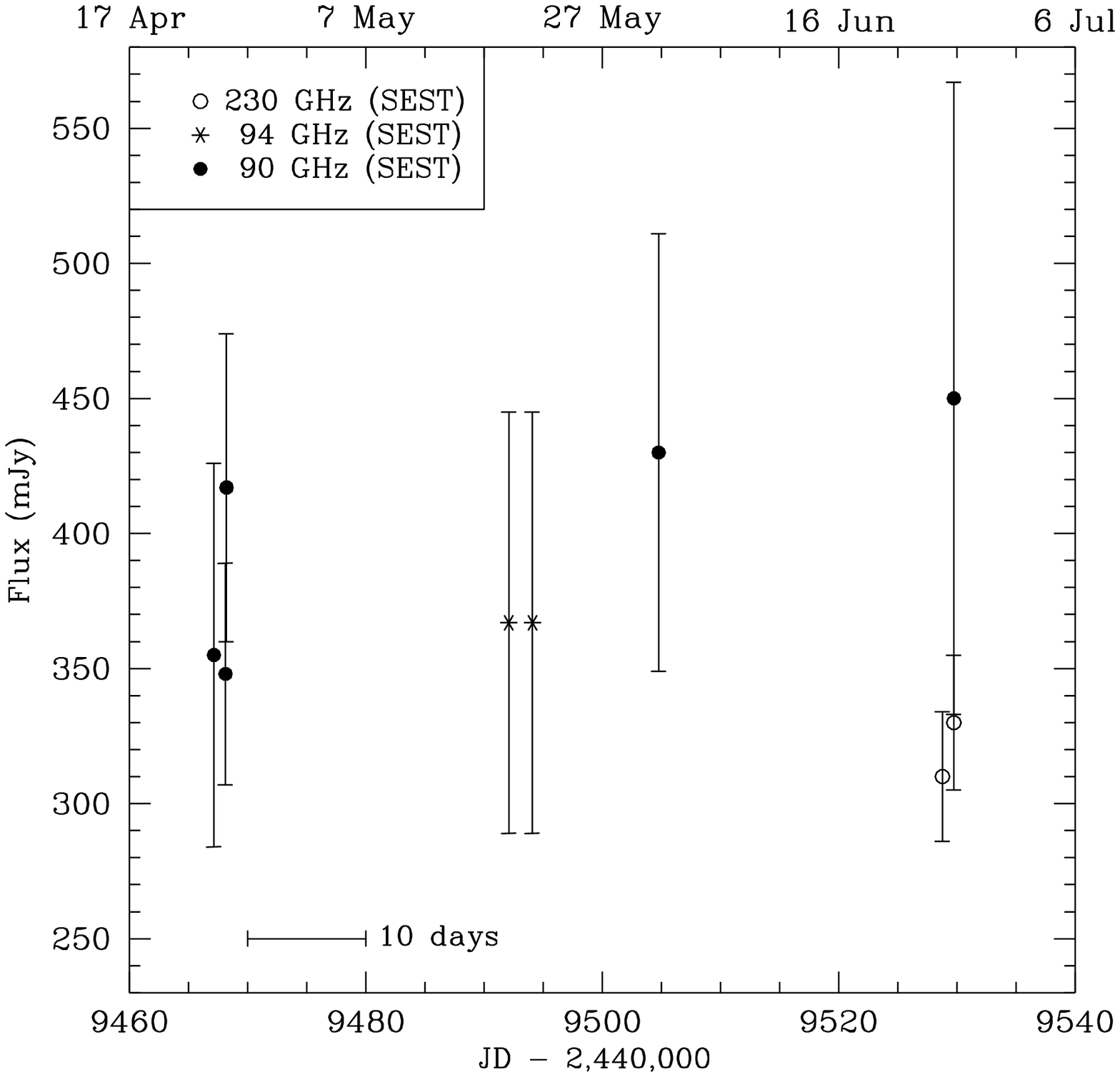,height=5in }}
\figcaption[fig3.ps]{The millimeter light curves from SEST. No variations are
obvious. A time scale bar is shown for comparison with the other figures.} 

\centerline{\psfig{figure=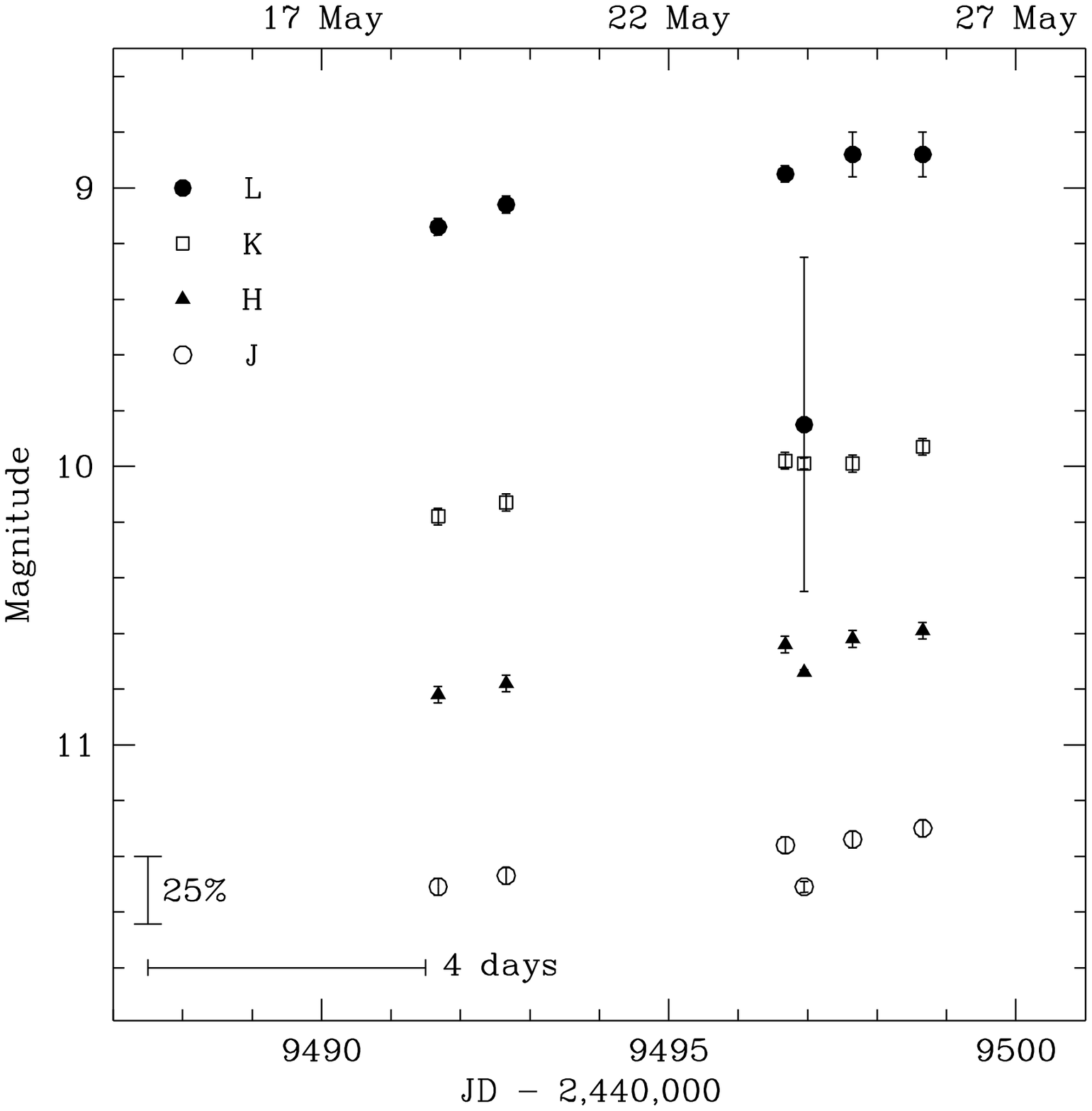,height=5in }}
\figcaption[fig4.ps]{The near-infrared light curves from 1994 May. All bands
increase by $\sim$0.3 mag over the period of observations. The $L$ magnitude 
on MJD 9496.95 is probably spurious, whereas the dip seen in $H$ and $J$ on 
the same day may be real but instrumental effects cannot be ruled out.
A time scale bar is shown for comparison with the other figures.} 

\centerline{\psfig{figure=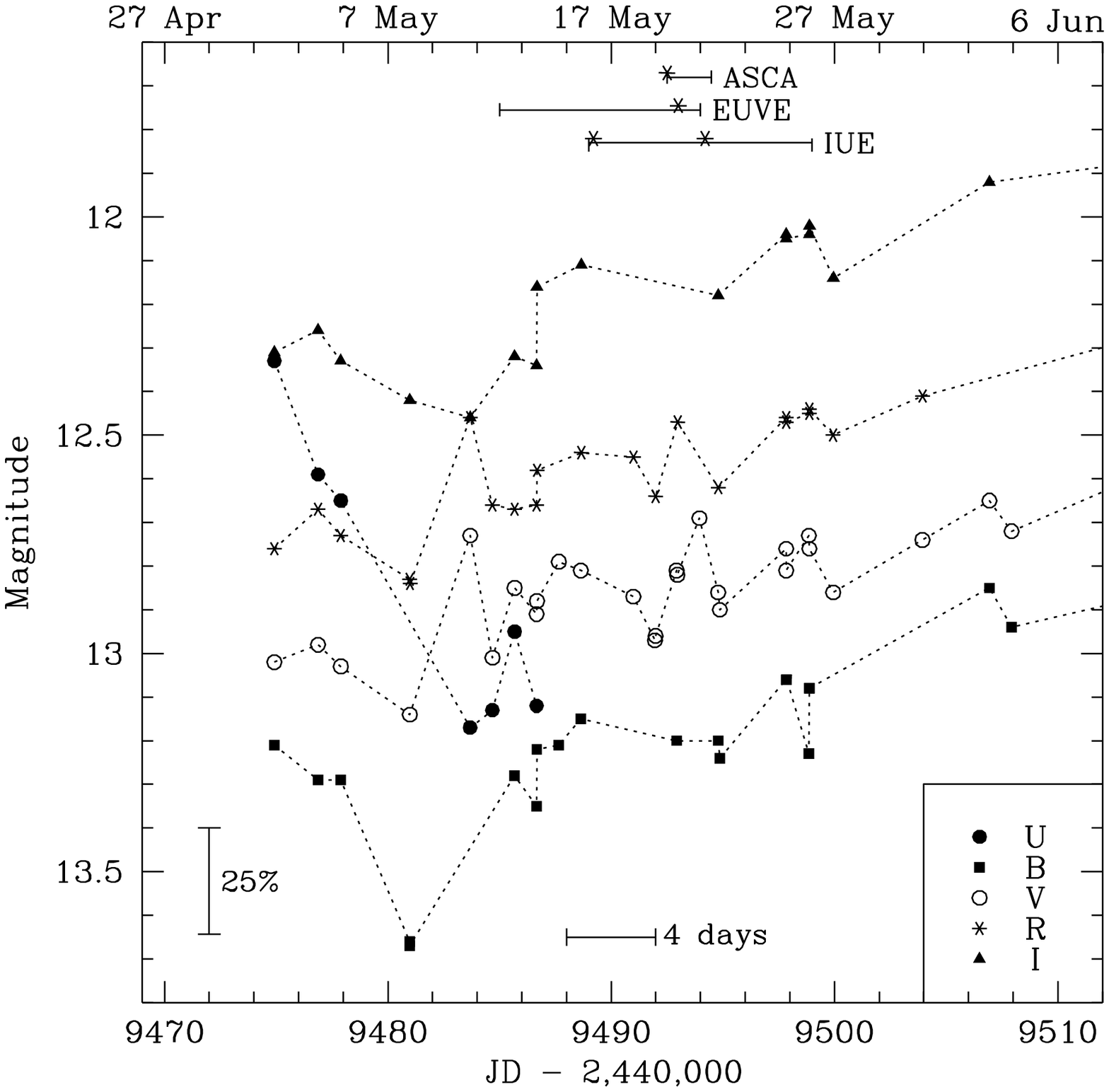,height=5in }}
\figcaption[fig5.ps]{The optical light curves during the multiwavelength
monitoring campaign, 1994 May. Where coverage is sufficient, it can
be seen that all bands vary
together. The $V$- and $R$-band fluxes show a feature between MJD 9492 and 9495
which corresponds to the UV flare. Note the $I$-band flux increase on 
$\sim$13 May (MJD 9486).  This corresponds to a variation of 0.01 mag 
min$^{-1}$. Uncertainties are the
size of the points or smaller (they range from $\lesssim$0.01 to $\sim$0.08
mag, with 0.01 mag being typical). The durations of the {\it ASCA, EUVE}, and
{\it IUE} experiments are shown at the top of the figure, and the midpoints of
the flares are indicated with asterisks.  A time scale bar is shown for
comparison with the other figures. The lines have been added to guide
the eye only.} 

\centerline{\psfig{figure=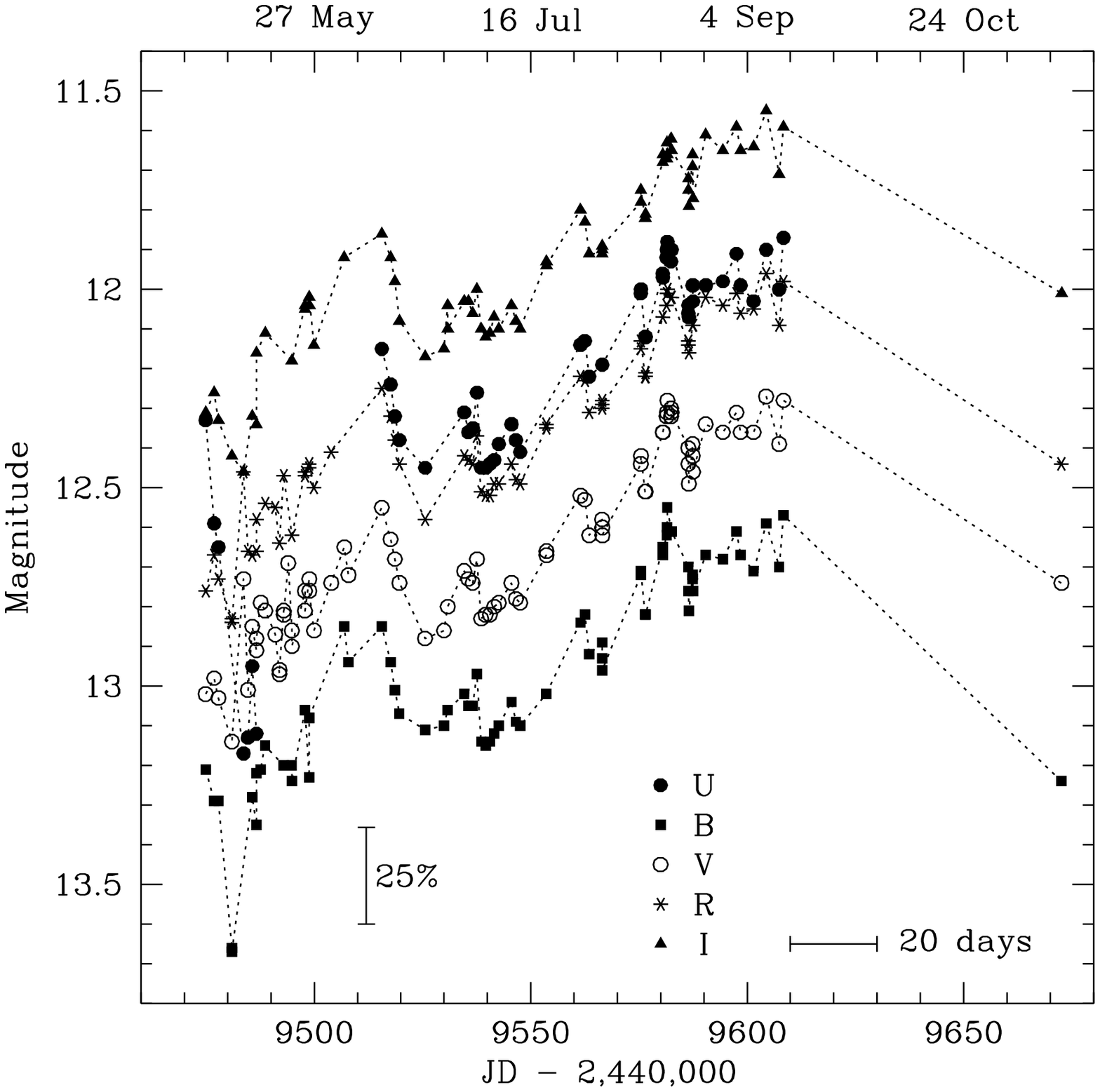,height=5in }}
\figcaption[fig6.ps]{The complete optical light curves from 1994 April to
November. Variations, from short-scale flickering
($\sim$0.2 mag in several days) to the longer-term trends, are of similar
amplitude at all wavebands with no measurable lags.  Uncertainties are the size of the
points or smaller (they range from $\lesssim$0.01 to $\sim$0.08 mag, with 0.01
mag being typical). A time scale bar is shown for comparison with the other
figures. The lines have been added to guide
the eye only.} 

\centerline{\psfig{figure=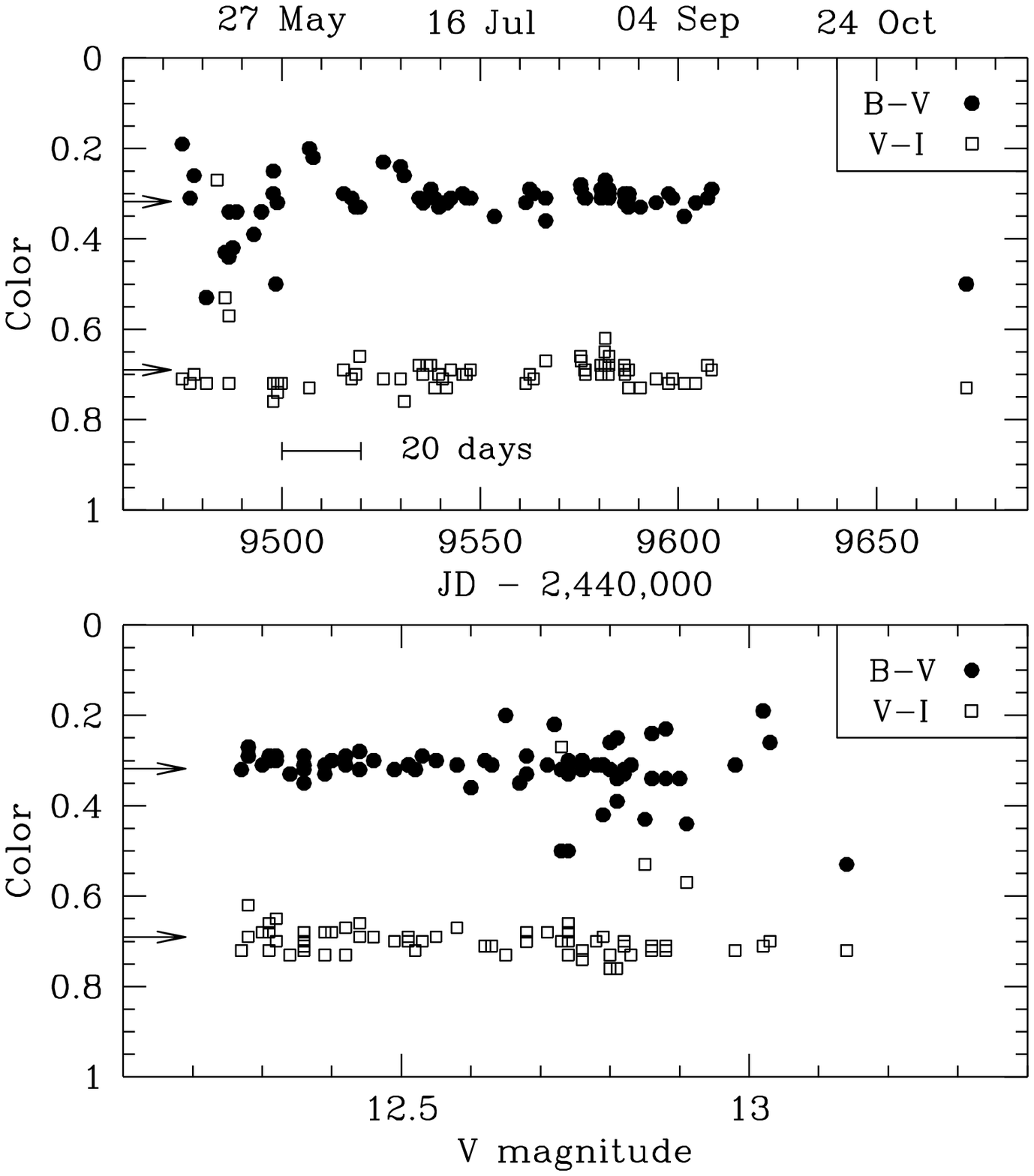,height=5in }}
\figcaption[fig7.ps]{The $B-V$ (solid circles) and $V-I$ (open squares) colors
for PKS 2155--304 as a function of time ({\it Top panel}) and $V$ magnitude
({\it Bottom panel}). The largest color variations occur when the source is
faint ($V \gtrsim 12.7$). Average colors are $\langle B-V \rangle = 0.32 \pm
0.02$ mag, $\langle V-I \rangle = 0.69 \pm 0.01$ mag, and are marked with the arrows.
The magnitudes used to calculate the colors were obtained simultaneously or
nearly so (within $\lesssim$ 10 minutes in most cases).} 

\centerline{\psfig{figure=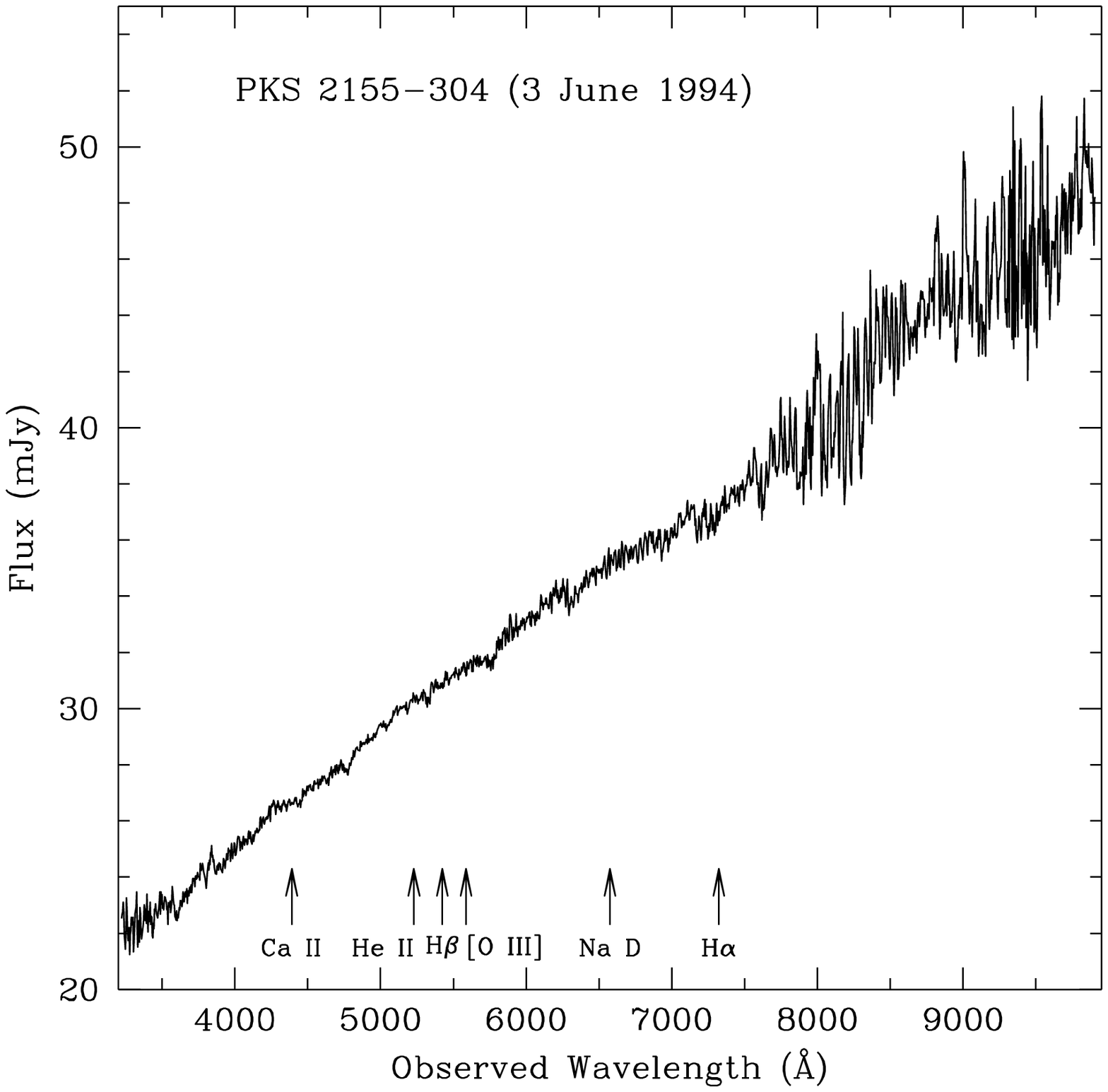,height=5in }}
\figcaption[fig8.ps]{The optical spectrum of PKS 2155--304 from Lick (MJD
9506.9875). A power law with index $\alpha = -0.71 \pm 0.02$ (where $F_{\nu} \propto
\nu^{\alpha}$) is a good representation of the spectrum, which is featureless
to an equivalent width limit of $\sim$1 \AA\ (or even 0.5 \AA\ in most places). 
Typical features, if present
at $z = 0.116$,
would be at the locations marked. The high frequency oscillations most
noticeable redward of 7500 \AA\ are produced by incompletely flattened CCD
interference fringes.} 

\centerline{\psfig{figure=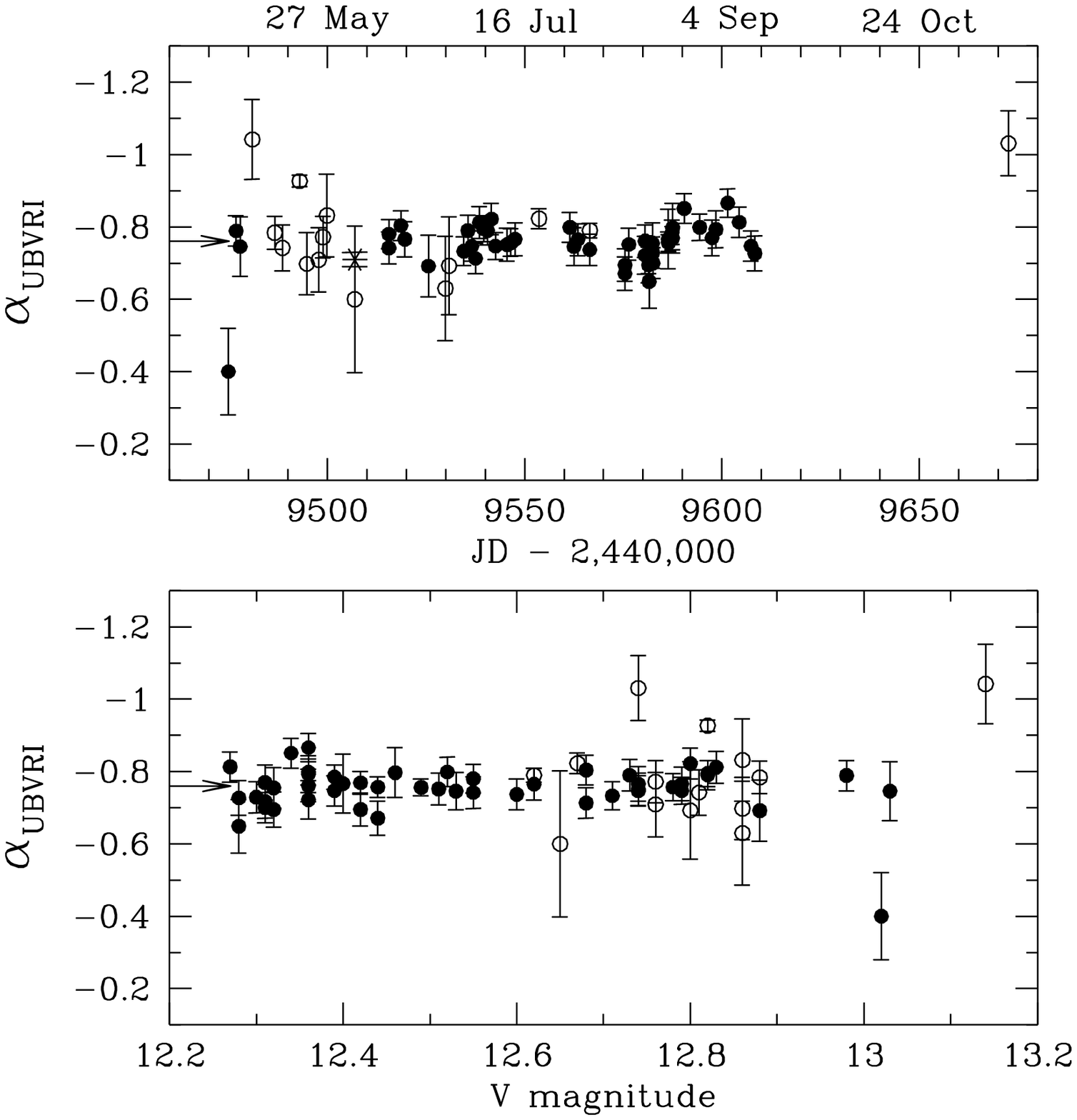,height=5in }}
\figcaption[fig9.ps]{{\it Top panel:} Slope of the total flux energy
distribution derived from fits ($F_{\nu} \propto \nu^{\alpha}$) to five
simultaneous {\sl UBVRI\/} measurements (filled circles) and three or four
simultaneous measurements (open circles). The average slope is $\langle
\alpha_{UBVRI} \rangle = - 0.76 \pm 0.03$ (arrow). The asterisk is the slope
from the Lick spectrum.  {\it Bottom panel:} Same as above, but as a function
of $V$-band magnitude.  There is a very slight steepening of the spectrum
with increasing magnitude, although this is not significant.} 

\centerline{\psfig{figure=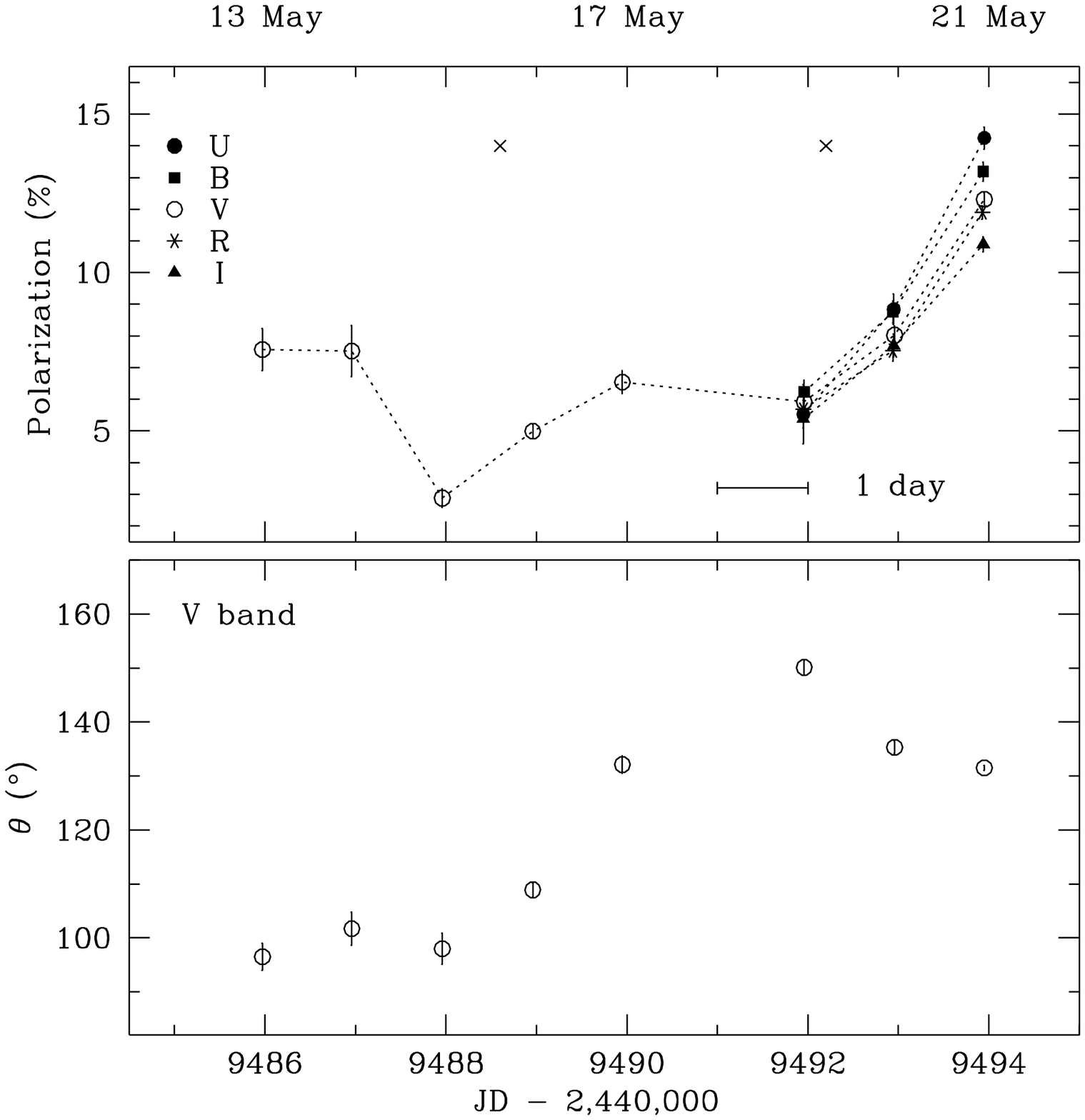,height=5in }}
\figcaption[fig10.ps]{{\it Top panel}: The polarization light curves from 1994
May (Mt. Lemmon, Arizona). The increase in polarization after MJD 9492 occurs
at all wavebands, and the wavelength dependent polarization is obvious. The
increases in polarization after MJD 9488 and 9492 occur at the same
time as the ultraviolet flaring events, the start times of which are marked
with an ``X'' (Urry \ea 1996). A time scale bar is shown for comparison with the
other figures. The lines have been added to guide the eye only.
{\it Bottom panel}: The polarization position angle for the $V$
band.  The preferred range is $\sim$90$^{\circ}$ - 150$\fdg$} 

\centerline{\psfig{figure=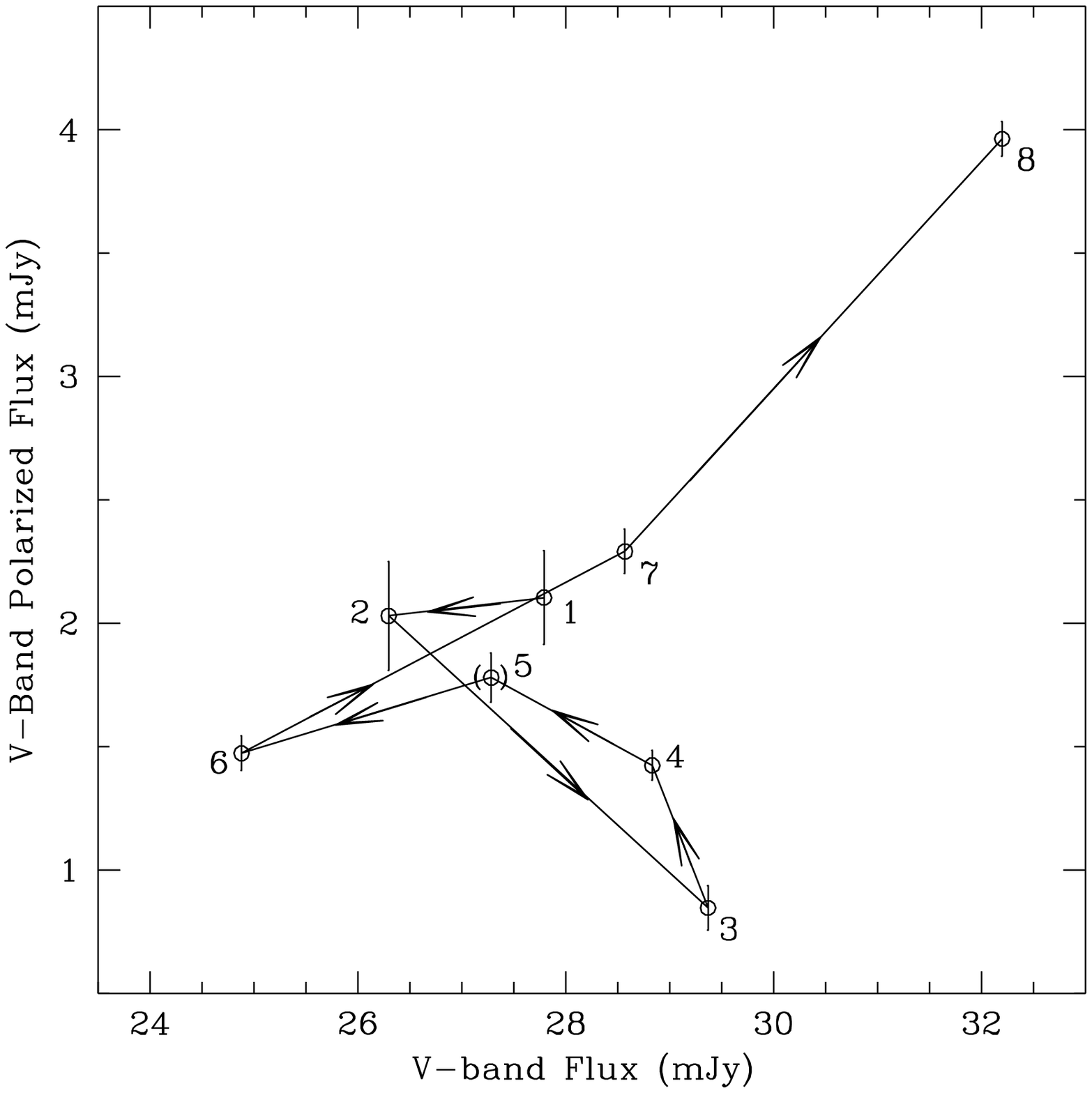,height=5in }}
\figcaption[fig11.ps]{The polarized $V$-band flux versus $V$-band flux for 1994
May, numbered in chronological order. PKS 2155--304 both brightens and fades
when the polarization increases.  The two ultraviolet flaring events occurred
after observations 3 and 6.} 

%\figcaption[fig12.ps]{{\it top panel}: The polarization light curves from 1994
%July to October (S\~ao Paulo). The source drops in polarization while
%increasing in brightness at all wavebands. {\it bottom panel}: The polarization
%position angle for the $V$-band data. Note the position angle is within the
%preferred range of $\sim$90$^{\circ}$ - 150$\fdg$} 

%\clearpage
%\clearpage
%\clearpage
%\clearpage
%\clearpage
%\clearpage
%\clearpage
%\clearpage
%\clearpage
%\clearpage
%\clearpage
%%\clearpage
%%\centerline{\psfig{figure=fig12.ps,height=8in }}

\end{document}